\documentclass[12pt]{article}%
\usepackage{cite}
\usepackage{heck}
\usepackage{graphicx}
\usepackage{multicol}
\usepackage{amsmath}
\usepackage{amssymb}
\usepackage[margin=1in]{geometry}

\providecommand{\U}[1]{\protect\rule{.1in}{.1in}}
\numberwithin{equation}{section}

\def\bea{\begin{eqnarray}}
\def\eea{\end{eqnarray}}

\newcommand{\scN}{\ensuremath{\mathcal{N}}}

\newcommand{\beq}{\begin{eqnarray}}
\newcommand{\eeq}{\end{eqnarray}}

\newcommand{\Gr}{\ensuremath{\mathop{\mathrm{Gr}}}}

\begin{document}

\date{November 2012}

\title{String Theory Origin of Bipartite SCFTs}

\institution{HarvardU}{\centerline{${}^{1}$Jefferson Physical Laboratory, Harvard University, Cambridge, MA 02138, USA}}

\institution{IAS}{\centerline{${}^{2}$School of Natural Sciences, Institute for Advanced Study, Princeton, NJ 08540, USA}}

\institution{PU}{\centerline{${}^{3}$Department of Physics, Princeton University, Princeton, NJ 08544, USA}}

\authors{Jonathan J. Heckman\worksat{\HarvardU}\footnote{e-mail: {\tt jheckman@physics.harvard.edu}}, Cumrun Vafa\worksat{\HarvardU}\footnote{e-mail: {\tt vafa@physics.harvard.edu}}, \\[4mm] Dan Xie\worksat{\IAS}\footnote{e-mail: {\tt dxie@ias.edu}} and Masahito Yamazaki\worksat{\PU}\footnote{e-mail: {\tt masahito@princeton.edu}}}

\abstract{We provide a string theory embedding for $\mathcal{N} = 1$ superconformal field
theories defined by bipartite graphs inscribed on a disk. We realize these theories
by exploiting the close connection with related $\mathcal{N} = 2$ generalized $(A_{k-1} , A_{n-1})$ Argyres-Douglas
theories. The $\mathcal{N} = 1$ theory is obtained from spacetime filling D5-branes wrapped on an algebraic curve
and NS5-branes wrapped on special Lagrangians of $\mathbb{C}^2$ which intersect
along the BPS flow lines of the corresponding $\mathcal{N} = 2$ Argyres-Douglas theory. Dualities of
the $\mathcal{N} = 1$ field theory follow from geometric deformations of the brane configuration
which leave the UV boundary conditions fixed. In particular we show how to recover the classification of
IR fixed points from cells of the totally non-negative Grassmannian
Gr$^{\rm tnn}_{k,n+k}$. Additionally, we present evidence that in the 3D theory
obtained from dimensional reduction on a circle, VEVs of line operators
given by D3-branes wrapped over faces of the bipartite graph
specify a coordinate system for Gr$^{\rm tnn}_{k,n+k}$.}

\maketitle

\enlargethispage{\baselineskip}

\setcounter{tocdepth}{2}
\tableofcontents

\newpage

\renewcommand\Large{\fontsize{16.5}{18}\selectfont}

\section{Introduction \label{sec:INTRO}}

A striking feature of string theory is the simple geometric characterization
it provides of diverse quantum field theories in various dimensions.
Non-trivial field theoretic dualities often have a simple geometric
underpinning in the string realization. Indeed, the rich mathematical
structure of string geometry often reveals itself in surprisingly diverse ways
in an effective field theory.

Recently a new class of $\mathcal{N}=1$ quantum field theories was discovered
which are specified by planar bipartite graphs inscribed on a disk \cite{Xie:2012mr}.\footnote{A
bicolored network or graph is a graph in which each vertex can be colored as black or
white. A bipartite graph or ``dimer'' refers to the case where each vertex only attaches to a vertex of the opposite
coloring. In the physical theory the bipartite and bicolored graphs define the same physical theory 
so we shall often use the two terms interchangeably.} This class of theories is related to, but distinct from, earlier
work on quiver gauge theories specified by a dimer on a torus
\cite{Hanany:2005ve,Franco:2005rj,Franco:2005sm,Feng:2005gw} (see
\cite{Kennaway:2007tq,Yamazaki:2008bt} for reviews). Given such a bipartite
graph, there is a corresponding quiver gauge theory with gauge groups, matter
fields and superpotential interaction terms specified by the connectivity of
the graph. While these theories
are non-conformal in the UV, it was conjectured in \cite{Xie:2012mr}
that they flow to non-trivial strongly coupled
IR fixed points. Moreover, thanks to the
mathematical results of \cite{Postnikov} for bicolored graphs,
these fixed points are classified by decorated
permutations, which in turn are specified by cells in the
totally non-negative Grassmannian Gr$^{\rm tnn}_{k,n+k}$.
Here the totally non-negative part (tnn) of the Grassmannian denotes the
subspace of Gr$_{k,n+k} \equiv GL_{k}(\mathbb{R}) \backslash$Mat$_{k,n+k}(\mathbb{R})$
with all Pl\"{u}cker coordinates non-negative. Closely related but distinct
bipartite theories have been considered in \cite{Franco:2012mm}.

The specification of a cell in the totally non-negative Grassmannian provides a simple way to
characterize possible Seiberg dual phases, that is, the equivalence class of
bipartite networks which define the \textit{same} IR fixed point.
This can be checked by appealing to the available
results in the mathematics literature, but it leaves open the underlying
question as to why these equivalences should hold, and more generally, the
physical origin of these bipartite graphs.

In this paper we provide an embedding in string theory for such
bipartite theories. Additionally, we show how lower order deformations
of the brane configuration translate to non-trivial field theoretic dualities, thus providing
a geometric characterization of Seiberg dual theories.
Moreover, we give a map between the coordinates of the Grassmannian
and operators of these $\mathcal{N}=1$ theories.

Our $\mathcal{N} = 1$ theories are realized by a configuration of
spacetime filling D5-branes and NS5-branes in type IIB\ string theory. The
background geometry is given by
$\mathbb{R}^{3,1}\times\mathbb{C}^{2}\times \mathbb{C}_{\bot}$, with all
branes sitting at the same point of
$\mathbb{C}_{\bot}$. In this geometry, we consider a stack of $N$ D5-branes wrapping the
algebraic curve $\Sigma$ defined as the zero set in $\mathbb{C}^2$ of:
\begin{equation}
y^{k}=x^{n}+\text{(deformations)}\ ,
\end{equation}
together with NS5-branes wrapping special Lagrangian (sLag) submanifolds of
$\mathbb{C}^{2}$ which intersect $\Sigma$ along one-dimensional subspaces.
This intersection pattern traces out a network of one-cycles on $\Sigma$ which partitions
the D5-brane into separate pieces, realizing the quiver gauge theories
for the top-dimensional cell of Gr$^{\rm tnn}_{k,n+k}$. We reach the
lower cells by brane recombination, corresponding to a partial Higgsing
of the quiver gauge theory.

Although this construction yields an $\mathcal{N}=1$ supersymmetric
theory, the geometric ingredients follow from an $\mathcal{N} = 2$
precursor.\footnote{A similar correspondence
has been studied for bipartite graphs on a torus \cite{Ooguri:2008yb}, where
the same geometry and the same quiver describe either a four-dimensional
$\mathcal{N}=1$ SCFT, or a one-dimensional quantum mechanics for a 1/2 BPS
particle inside a four-dimensional $\mathcal{N}=2$ theory.} For example, the
curve $\Sigma$ can also be wrapped by an M5-brane, realizing a 4D $\mathcal{N} = 2$ theory. In the limit
where all deformations are switched off, this produces the $\mathcal{N} = 2$ generalized $(A_{k-1} , A_{n-1})$
Argyres-Douglas fixed point (see for example Figure \ref{fig.G2flow} for the comparison between 
4d $\mathcal{N}=1$ theory and $\mathcal{N}=2$ theory). Once the deformations of the geometry are switched on,
the one-cycles of the intersection correspond to a network of BPS flows
which preserve a fixed $\mathcal{N}=1$ subalgebra of the $\mathcal{N}= 2$ supersymmetry algebra.
Given such a one-dimensional network, there exists a unique set of special
Lagrangians in $\mathbb{C}^{2}$ which intersect $\Sigma$, realizing our quiver gauge theories.

The geometric characterization of these quiver gauge theories also allows us
to delineate possible Seiberg dual phases. Fixing the boundary conditions of
all of the branes, lower order deformations of the algebraic
curve $\Sigma$ translate to possible jumps in the connectivity of the BPS flow lines. In
the $\mathcal{N}=1$ quiver gauge theory this corresponds to Seiberg duality.

The string construction also leads us to a conjectural relation between
the coordinates of Gr$^{\rm tnn}_{k,n+k}$ and the VEVs of line operators in the dimensional
reduction of the 4D theory. The basic correspondence involves D3-branes suspended
between NS5-branes which are wrapped over two directions of the 4D spacetime.
Compactifying the 4D spacetime on the geometry $S_{(t)}^{1}\times MC_{q}$ with
$MC_{q}$ a Melvin cigar geometry, we show that in the small $S_{(t)}^{1}$ limit,
these operators are closely related to line operators of the $\mathcal{N}=2$ Argyres-Douglas theory.
Quite remarkably, this class of line operators specify coordinates of the Grassmannian.

The rest of this paper is organized as follows. In section \ref{sec:REVIEW} we
review the defining elements of the bipartite quiver gauge theories and
the correspondence with the totally non-negative Grassmannian. In section \ref{sec:TOP}
we discuss in more detail the theories defined by the top cell of
Gr$^{\rm tnn}_{k,n+k}$, since all
other fixed points follow from Higgsing of this case. In preparation for
our discussion of $\mathcal{N} = 1$ SCFTs, in section \ref{sec:N=2} we review
some aspects of BPS quivers for $\mathcal{N} = 2$ theories, and in particular the close
connection between the bipartite graphs for Gr$^{\rm tnn}_{k,n+k}$ and the generalized $(A_{k-1} , A_{n-1})$
Argyres-Douglas theories. In section \ref{sec:STRING} we present our proposal for a string theory
realization of these quiver gauge theories. Moreover, we show that this
construction automatically explains the fact that various gauge theories defined by related
bipartite networks flow to the same IR fixed point. In section \ref{sec:IR} we explain how in the compactification down to three dimensions,
line operators of the theory specify coordinates of the Grassmannian. Section \ref{sec:CONC} presents our
conclusions and possible directions of future investigation. Some additional
technical details are deferred to the Appendices.

\section{Bipartite $\mathcal{N}=1$ SCFTs \label{sec:REVIEW}}

In this section we introduce a class of quiver gauge theories defined by planar
bipartite networks \cite{Xie:2012mr}. We first review the basic properties of
these theories and then explain how the combinatorics of the
graphs can be related to a cell decomposition of the totally non-negative Grassmannian.
This provides a powerful way to characterize possible IR\ fixed points of
the quiver gauge theories. One of our tasks in this paper will be to give a
simple explanation for these relations from a brane construction in string theory.

\subsection{Definitions}
To frame the discussion to follow we begin by defining in more precise terms
our bipartite theories. A bicolored network (or ``graph'') is
a graph in which the vertices can be colored as either black
or white. A bipartite graph or dimer refers to a further specialization where only black vertices are adjacent to white
vertices and visa versa. Throughout this paper we restrict attention to graphs
which are connected and can be drawn in the plane, and in which all legs
which attach to only one vertex are external, that is, they can be drawn as
ending on a circle at infinity which surrounds the planar graph.

\begin{figure}[t!]
\centering{\includegraphics[scale=0.48]{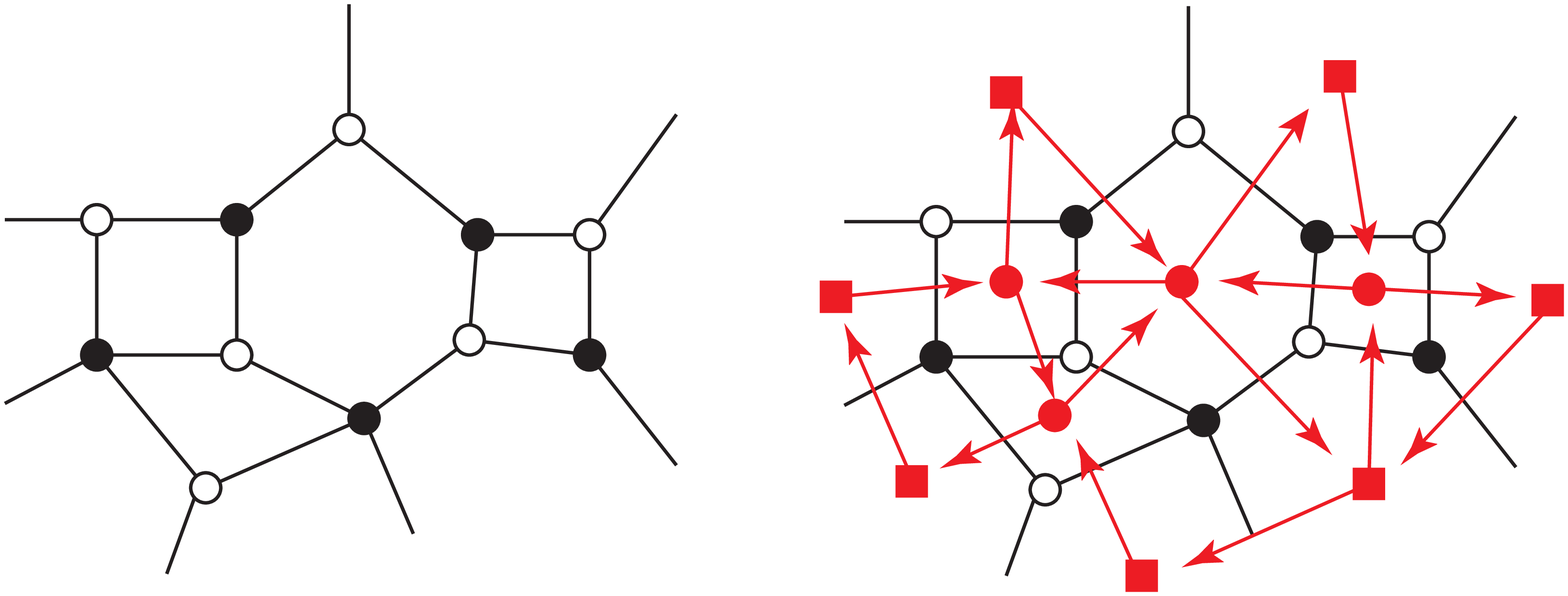}}
\caption{Depiction of a planar bicolored graph, corresponding to a cell
 of $\Gr^{\rm tnn}_{3,7}$.
The dual graph defines a 4D $\mathcal{N} = 1$ quiver gauge theory. In the original
bicolored graph, each face specifies either a gauge group (red circle) or a flavor group (red square). Each
interior edge specifies an oriented bifundamental, and each interior vertex defines a superpotential term. For exterior
edges and vertices there is only a contribution from black vertices.}
\label{networkeg}
\end{figure}

Given such a bicolored graph we associate a 4D $\mathcal{N}=1$ quiver gauge theory
to it by taking the graph dual. In more detail:

1. Attach a $U(N)$ gauge group to each closed, i.e. compact face and a $U(N)$
flavor group to an open, i.e. non-compact face on the boundary.\footnote{We could
consider more general theories by choosing different ranks for different
faces, as long as they are anomaly free.} These are represented in figure \ref{networkeg}
by red circles and squares, respectively.

2. Each edge gives a bifundamental field. The chirality of the
bifundamental, or equivalently the orientation of the arrow of the quiver
diagram, is determined by the black/white coloring of the vertices; the quiver
arrows form a clockwise (resp. counterclockwise) loop around the black (resp. white)
vertices. There is an important exception to this rule: we do not associate a
bifundamental matter to external edges which attach to a white vertex.

3. There is a superpotential term for each vertex, given by the trace of the
product of all the bifundamentals around the vertex.

Note that in this construction, all of the gauge symmetries,
as well as flavor symmetries, are automatically anomaly free.
This is because there are an equal number of ingoing and
outgoing arrows.

The rules above are chosen such that the mathematical result of
\cite{Postnikov} can be directly translated into physical statements about the IR
fixed points of the 4D $\mathcal{N}=1$ theories. The rules mostly follow the string theory
literature on dimers on a torus \cite{Hanany:2005ve,Franco:2005rj}. However as we already mentioned, there
is one crucial difference due to the fact that white vertices attached to
an external leg are treated differently. Note that the paper of
\cite{Franco:2012mm} does not include this rule, and hence our theory
is different from the theory defined there. Our definition is
chosen so that the square move of the bipartite graph corresponds to
Seiberg duality of the 4D $\mathcal{N}=1$ theory, even for faces which are
adjacent to flavor branes.

One of our aims in this paper will be to explain
how these seemingly ad hoc conditions naturally arise in a string construction
(see in particular the discussion around Figure \ref{floweg}).

\subsection{IR\ Equivalence Classes and the Grassmannian}

Having introduced our quiver gauge theories, it is natural to study
their properties in the infrared. In the remainder of this section we explain
how the geometry of the Grassmannian naturally arises in a classification of
IR fixed points.

\begin{figure}[t!]
\centering{\includegraphics[scale=0.4]{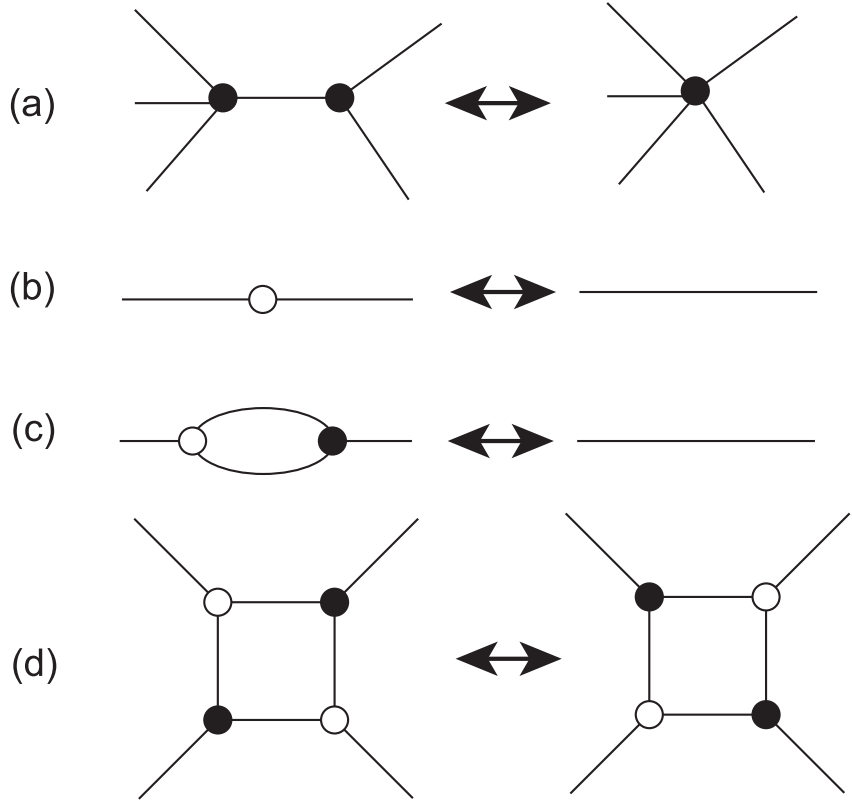}}
\caption{Depiction of graphical operations on a planar bicolored network which
preserve the IR fixed point of the associated 4D $\scN=1$ theory.
(a,b): integrating out massive matter fields,
 (c): confinement of a gauge group,
(d): Seiberg duality.
}
\label{fig.moves}
\end{figure}

Given an $\mathcal{N}=1$ theory defined from a bicolored network,
we want to determine the graphical operations which would not change the IR physics.
First of all, one could integrate out massive fields in the theory,
and such operations are mapped to
merging same-colored vertices connected by edges (figure
\ref{fig.moves}(a)) or
removing the degree two vertices (figure \ref{fig.moves}(b)).
We can also integrate out the IR-confining
gauge group with $N_f=N_c$,
and this operation is mapped graphically to the so-called bubble
reduction (figure \ref{fig.moves}(c)).

Far more non-trivial is the Seiberg duality acting on a gauge group in the conformal window.
In our case, this only happens for square faces corresponding to gauge groups with $N_f=2N$.
Seiberg duality then corresponds to the so-called
\textquotedblleft square move\textquotedblright\ (figure \ref{fig.moves}(d)).

It is very useful to restrict to bicolored networks which do not have bubbles
under any of the moves introduced earlier. Such a
\textquotedblleft reduced\textquotedblright\  graph contains the minimal UV data necessary to characterize the IR fixed point.
There is a powerful combinatorial algorithm, based on ``zig-zag paths'', for determining whether or not
a network is reduced \cite{Xie:2012mr}. In the following we
will restrict to the case where such a reduction has already been performed.

Given two reduced graphs, we can now ask whether these
two theories will flow to the same IR\ fixed point.
This can also be answered with the help of zig-zag paths,
and permutations defined from them.
Namely two reduced networks define the same IR SCFT if they define the same
permutation.

\begin{figure}[t!]
\small
\centering
\includegraphics[scale=0.48]{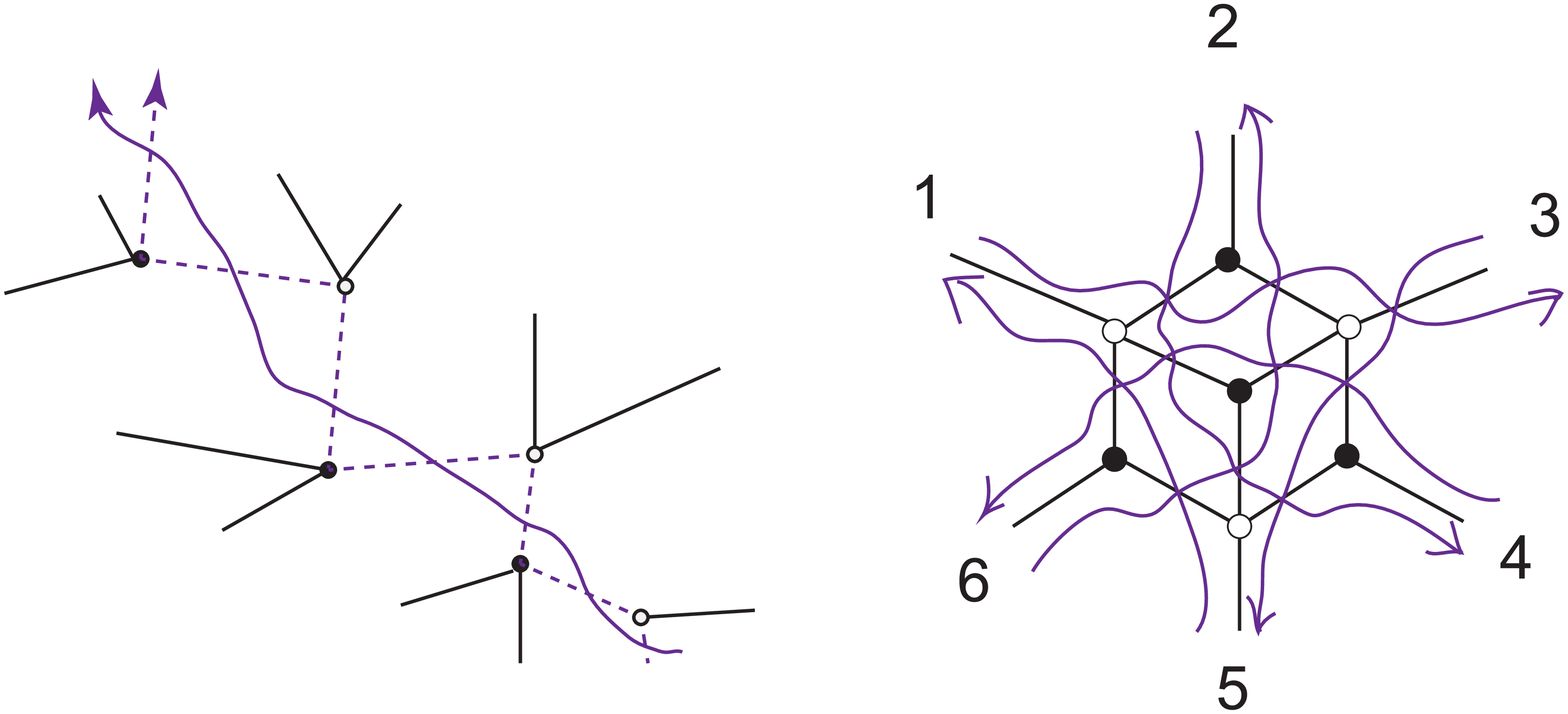}
\caption{An example of a zig-zag path on a bipartite graph. The rule is specified by following
a given leg, and turning left at a white vertex and right at a black
 vertex (left). For a planar bipartite graph,
this yields a permutation assignment between one external leg and
 another (right).}
\label{zig}
\end{figure}

To define a zig-zag path (or an associated
\textquotedblleft strand\textquotedblright) and a permutation,
let us label all of the
external legs according to a counter-clockwise cyclic ordering on the
boundary circle ``at infinity''.
Starting from a given external leg (say the $i$-th leg) at infinity and following
it into the interior region of the graph, we obtain a path on the graph
by turning left at the white vertices and right at the black
vertices, finally ending at another external leg, say the $j$-th leg (figure \ref{zig}).
Each zig-zag path around the graph defines a permutation $i \rightarrow \pi(i):=j$
of $\mathfrak{S}_{n+k}$, where we have denoted the number of external legs by
$n+k$, to match with the notation later.

There is a remarkable correspondence between this
permutation and the cell decomposition of a totally non-negative Grassmannian
$\Gr^{\rm tnn}_{k,n+k}$ \cite{Postnikov}.\footnote{
Each cell is a ball, and is roughly a subspace of the real
Grassmannian with some of the Pl\"{u}cker
coordinates positive, with the rest set to zero.}
This means that the IR fixed points of our $\mathcal{N}=1$ theories
can also be uniquely labeled by the cells of the $\Gr^{\rm tnn}_{k,n+k}$.
Here the values of $k$ and $n$ can be read off from the formula
\begin{equation}
k-n=\sum_{\rm black}({\rm deg}(v)-2)-\sum_{\rm white}({\rm deg}(v)-2) \ ,
\label{kn}
\end{equation}
where ${\rm deg}(v)$ denotes the number of edges attached to a
vertex.\footnote{Let us note that although there is a geometric
duality which interchanges $k\leftrightarrow n$, these graphs define different
IR fixed points.}

The details of this correspondence are explained in Appendix \ref{sec:measurement}.
In particular we explain how coordinates of the Grassmannian are mapped
to the weights of faces in the bicolored network. In section \ref{sec:IR} we
give a physical interpretation for these parameters.

\section{The Top Cell of Gr$^{\rm tnn}_{k,n+k}$ \label{sec:TOP}}

In this section we discuss in more detail the theories defined by the top-dimensional cell of the
totally non-negative Grassmannian Gr$^{\rm tnn}_{k,n+k}$.
These theories are canonical in the sense that a theory
associated with a lower cell (corresponding to going to the boundary of the top cell) can
be obtained from partial Higgsing of the top cell theory.

\subsection{The Quivers}

\begin{figure}[t!]
\small
\centering
\includegraphics[width=\linewidth]{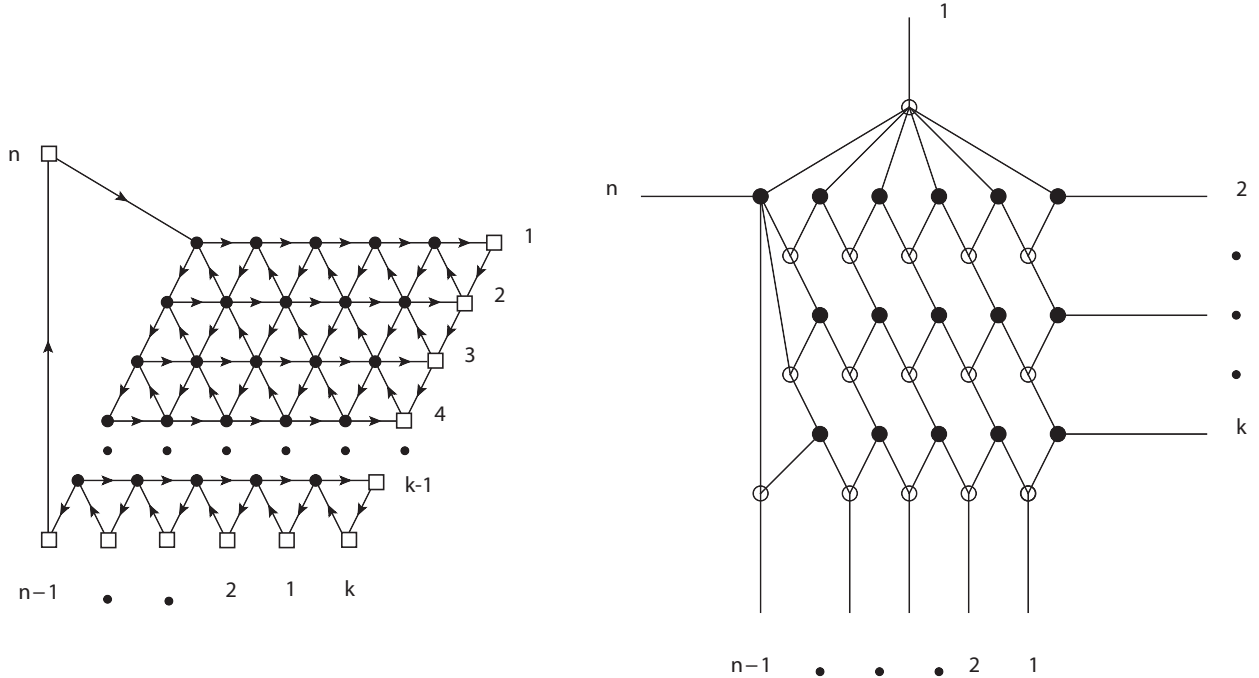}
\caption{The quiver (left) and the bipartite graph (right) associated with the
 top-dimensional cell of $\Gr^{\rm tnn}_{k,n+k}$. Square nodes
 of the quiver diagram denote flavor group factors and correspond to non-compact
 faces of the bipartite graph.}
\label{higher}
\end{figure}

The top-dimensional cell of $\Gr^{\rm tnn}_{k,n+k}$
is a cell with all the Pl\"{u}cker coordinates positive.
This corresponds to a permutation
\begin{equation}
\pi_{\mathrm{top}}(i)=i+k\quad(\text{mod}\,\,n+k),\quad i=1,\ldots,n+k \
 ,
\label{topperm}
\end{equation}
and bicolored graph and corresponding quiver shown in figure \ref{higher}.
We can construct these graphs from \eqref{topperm}
by use of the algorithm explained in \cite{Xie:2012mr, Postnikov}.
One can check explicitly that the permutation defined from the zig-zag
paths is indeed given by \eqref{topperm}.
The face of the network is either a square or hexagon.
The total number of $U(N)$ gauge groups is $(n-1)(k-1)$, and the number of flavor groups
is $n+k$.

Let us note that the geometric operation which interchanges $k$ and $n$ in the Grassmannian corresponds to
switching all black and white vertices. This leaves the connectivity of the internal parts of the quiver
unchanged, though it will alter the number of gauge singlets, as we treat external
legs attached to black and white vertices differently.

Finally, in the special case where $k = 2$, we have a single row of squares with alternating black and
white vertices. Each white vertex supports an external leg, but only the leftmost and rightmost black vertices support an
external leg. According to our rules specified in section \ref{sec:REVIEW}, this means there are
two gauge singlets. We obtain a distinct quiver gauge theory by interchanging $k$ and $n$. In
the graph, this corresponds to switching all black and white vertices, so that there are $n$ gauge
singlets for the top cell of Gr$^{\rm tnn}_{n,n+2}$.

\subsection{Infrared R-Symmetry \label{ssec:RCHARGES}}

In this subsection we discuss some properties of the infrared fixed points for the theories
associated with the top cell of Gr$_{k,n+k}^{\text{tnn}}$. We begin with some general considerations
which apply to all of our bipartite theories. We then present evidence for a
surprisingly rich phase structure as a function of $k$ and $n$.

In supersymmetric theories, a wealth of information is encoded in the infrared
R-symmetry. This is given by a linear combination of the UV R-symmetry and all
abelian flavor symmetries of the IR theory:%
\begin{equation}
R_{IR}=R_{UV}+\sum_{i}t_{i}F_{i}\ .
\end{equation}
The list of such abelian flavor symmetries includes all of the central
$U(1)\subset U(N)$ factors of the UV flavor symmetries, but can also include
emergent $U(1)$ symmetries. In general, it can be difficult to determine when
such accidental $U(1)$'s appear. One clear indication of their presence is
signalled by the existence of an otherwise consistent looking fixed point
which nevertheless naively contains an operator with dimension below the
unitarity bound. Assuming one has taken into account all such $U(1)$'s, the
procedure of $a$-maximization amounts to maximizing a trial $a$-function
\cite{Intriligator:2003jj}:%
\begin{equation}
a_{\mathrm{trial}}[t]=\frac{3}{32}\left(  3\,\text{Tr}R_{IR}^{3}%
-\text{Tr}R_{IR}\right)
\end{equation}
over the $t_{i}$, subject to all constraints imposed by a consistent IR\ fixed point.

Consider now the procedure of $a$-maximization for our bipartite theories. The analysis
is quite similar to the case of torus dimers discussed in \cite{Franco:2005rj, Franco:2005sm}.
Since we have a weakly coupled UV definition of the theory, we can appeal to anomaly
matching considerations to work directly in terms of the UV\ basis of fields.
In a general bipartite theory the constraints on the coefficients $t_{i}$ are
specified by two types of conditions. Letting $R(X)$ denote the R-charge of a
bifundamental $X$, the vanishing of the NSVZ beta function for each
face implies (see e.g. \cite{Franco:2005rj, Franco:2005sm}):%
\begin{equation}
\text{For each face }F: \underset{X\text{ on }F}{\sum}(1 - R(X) )= 2\ , \label{faceconstraint}%
\end{equation}
where the sum over \textquotedblleft$X$ on $F$\textquotedblright\ includes
all bifundamentals attached to the face $F$. Even in the absence of
the superpotential interaction terms, this will often produce a non-trivial
fixed point. Adding the superpotential deformations can be viewed as a further
deformation of the theory. In order for this deformation to be a
relevant/marginal deformation, it must have R-charge $+2$ in the IR:%
\begin{equation}
\text{For each vertex }V:\underset{X\text{ on }V}{\sum}R\left(  X\right)
=2\ . \label{vertexconstraint}%
\end{equation}
The procedure of $a$-maximization then corresponds to varying with respect to
these trial R-charges, subject to the constraints of equations
(\ref{faceconstraint}) and (\ref{vertexconstraint}). We note that in contrast to the
case of dimer models on a torus, there can be gauge singlets which participate in only a
single superpotential term. When such a gauge singlet multiplies a meson, this meson will
vanish in the chiral ring. In other words, such a meson does not parameterize the vacuum
structure of the low energy theory.

\begin{figure}[t!]
\centering\includegraphics[scale=1.1]{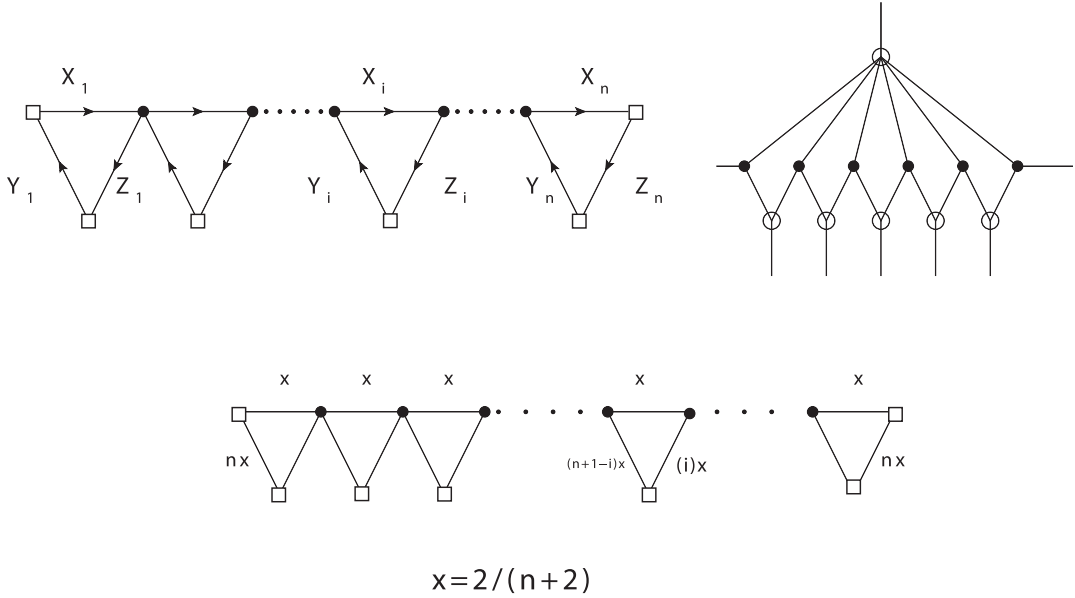}
\caption{The quiver (left), the bipartite graph
(right) and the ``naive'' R-charge assignments (bottom) for the top-dimensional cell of
Gr$^\mathrm{tnn}_{2,n+2}$. Square nodes denote flavor group factors. All of the
R-charge assignments of bifundamentals are proportional to $x = 2/(n+2)$.}%
\label{canonical}%
\end{figure}

Let us consider in more detail the analysis of R-charge assignments for the
$k=2$ theories, i.e., for the top cell of Gr$_{2,n+2}^{\text{tnn}}$. Now, as
we have already mentioned, one of the key points of the classification by
cells of the Grassmannian is that different bipartite graphs may nevertheless
flow to the same IR fixed point. Since the flavor symmetries remain the same
after Seiberg duality, it is enough to analyze the R-charge assignments in any
given chamber. A simple choice of duality frame is the one depicted in figure
\ref{canonical} which consists of a single row of squares (i.e.
\textquotedblleft diamonds\textquotedblright) which share a common vertex.

The flavor symmetries of our system include all of the central $U(1)$ factors
of the non-compact flavor faces. This provides $n+2$ flavor symmetries.
Additionally, there are $n-1$ \textquotedblleft baryonic\textquotedblright%
\ flavor symmetries corresponding to the central $U(1)$ attached to each gauge
group factor. In a theory with gauge group $SU(N)$, these can play the role of
flavor symmetries for dibaryon operators of the form:%
\begin{equation}
\mathcal{B}_{G_{i}G_{j}}=\varepsilon_{\overline{j_{1}},...,\overline{j_{N}}%
}\varepsilon^{i_{1}...i_{N}}Q_{i_{1}}^{\overline{j_{1}}} ... Q_{i_{N}}^{\overline{j_{N}}} \ ,%
\end{equation}
constructed from a bifundamental $Q$ between groups $G_{i}$ and $G_{j}$. This
can also happen in our case, because the central $U(1)$ of a gauge group
factor decouples in the infrared.

In addition to these UV symmetries, there can also be accidental $U(1)$
symmetries in the infrared theory. We will present some evidence that a
self-consistent treatment of these fixed points involves the
inclusion of such accidental $U(1)$'s. To argue for their existence, we
first proceed as if they did not exist.

In the duality frame of figure \ref{canonical}, it is convenient to label the
bifundamentals according to small triangles formed in the quiver. For
$i=1,...,n$, we denote by $X_{i}$ the horizontal arrows, $Y_{i}$ the arrows
pointing up and to the left, and $Z_{i}$ the arrows pointing down and to the
left. For notational simplicity, we shall often write $X_{0}\equiv Z_{1}$
interchangeably. We denote by lowercase letters the corresponding R-charges.
Owing to the $\mathbb{Z}_{2}$ symmetry of the quiver, we have:%
\begin{equation}
x_{i}=x_{n+1-i}\text{, \ \ }y_{i}=z_{n+1-i}\text{, \ \ and \ \ }%
z_{i}=y_{n+1-i} \ .
\end{equation}
The trial $a$-function is given by:%
\begin{equation}
a_{\mathrm{trial}}=\frac{3N^{2}}{32}\left(  2(n-1)+\underset{i=1}{\overset
{n}{\sum}}\left(  f\left(  x_{i}\right)  +f(y_{i})+f(z_{i})\right)
				    \right) \ ,
\end{equation}
where $f(x)\equiv3(x-1)^{3}-(x-1)$. The overall constant is due to the
contribution from the gauge group factors, and the factor of $N^{2}$ is due to
the fact that each field in the UV is an $N\times N$ matrix. To perform
$a$-maximization, we first impose the beta function and superpotential
conditions of equations (\ref{faceconstraint}) and (\ref{vertexconstraint}),
respectively:%
\begin{equation}
\begin{split}
x_{i+1}+x_{i}+y_{i+1}+z_{i}  &  =2 \ ,\\
x_{i}+y_{i}+z_{i}  &  =2 \ ,
\end{split}
\end{equation}
for $i=1,...,n$. Here, the lowercase letter denotes the R-charge of the field.
Solving in terms of the $x_{i}$ yields:%
\begin{equation}
y_{i}=2-\underset{j=0}{\overset{i}{\sum}}x_{j}\quad \text{ and }\quad z_{i}=\underset
{j=0}{\overset{i-1}{\sum}}x_{j} \ .
\end{equation}
We therefore need to perform $a$-maximization over the $n+1$ independent
parameters $x_{0},...,x_{n}$. To simplify $a_{\mathrm{trial}}\left[  x\right]
$, we use the fact that $f(2-x)=-f(x)$. After some algebraic manipulation, one
finds that the trial $a$-function considerably simplifies to:%
\begin{equation}
a_{\mathrm{trial}}\left[  x\right]  =\frac{3N^{2}}{32}\left(  2(n-1)-f\left(
x_{\Sigma}\right)  +\underset{i=0}{\overset{n}{\sum}}f(x_{i})\right) \ ,
\label{atrialfirst}%
\end{equation}
where we have introduced $x_{\Sigma}\equiv x_{0}+...+x_{n}$. The local maximum has all $x_{i}$ equal,
and given by:
\begin{equation}\label{naiveR}
x_{i}=\frac{2}{n+2} \ .
\end{equation}

An important subtlety with this analysis is that it does not take into account the possibility of
accidental $U(1)$ symmetries in the infrared. That such symmetries will appear
can be seen from examining the R-charge assignments for composite operators which
we refer to as mesons:
\begin{align}
\mathcal{M}_{j}  &  =X_{1}...X_{j}Z_{j} \ ,\label{meson}\\
\mathcal{M}_{j}^{\prime}  &  =Y_{n-j+1}X_{n-j+1}...X_{n} \ , \label{mprime}%
\end{align}
Using the R-charge assignments of equation (\ref{naiveR}), one might conclude that
the $j^{th}$ meson has R-charge $4j/(n+2)$. In other words, for $j$ sufficiently small
and $n$ sufficiently large, there can be an apparent violation of the unitarity bound.
Now, although the mesons $\mathcal{M}_{1}$ and $\mathcal{M}_{1}^{\prime}$ have R-charge
below $2/3$ when $n > 4$, we observe that in the chiral ring they are set to zero, so there is
no issue with a violation of the unitarity bound for these operators. By contrast,
for $j \geq 2$, there could indeed be an apparent violation of the unitarity bound. This does \textit{not}
signal an inconsistency of the theory, but rather, provides a strong indication
for an accidental $U(1)$ in the infrared.

\begin{figure}[t!]
{\small \centering
\includegraphics[width=\linewidth]{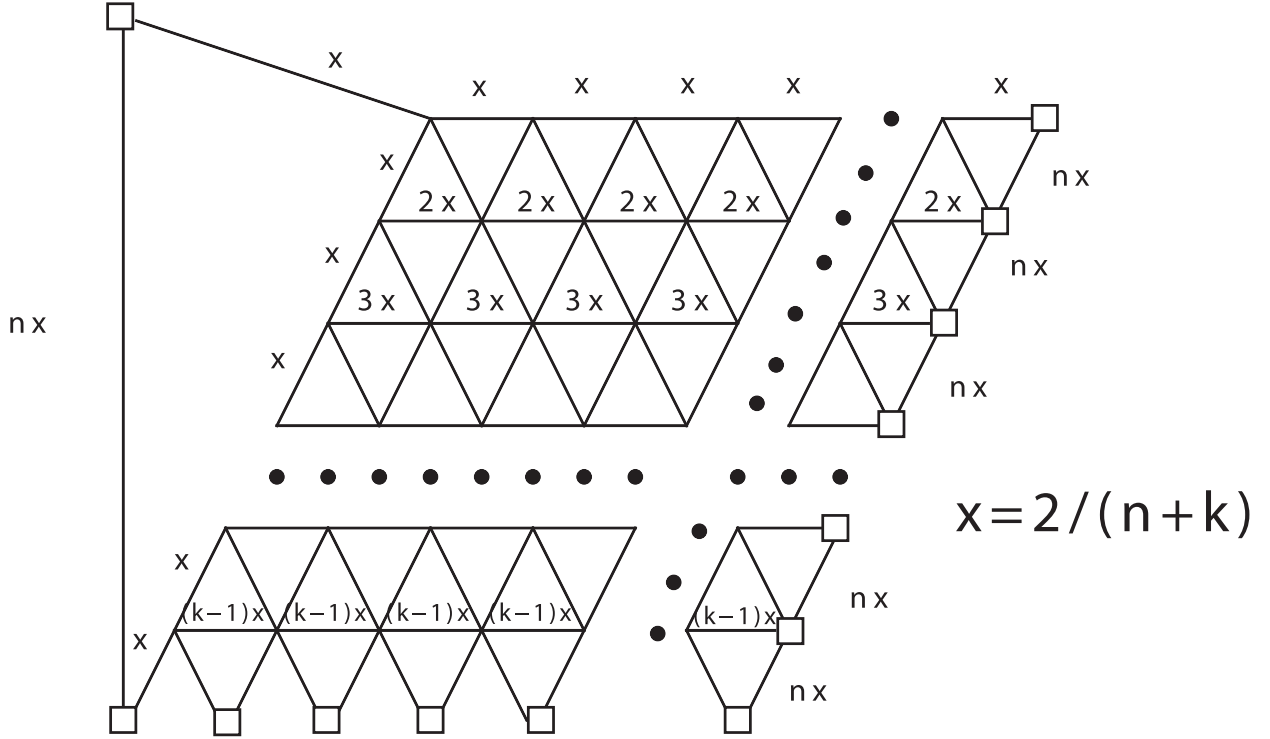} }
\caption{Applying $a$-maximization without taking
into account the decoupling of mesons, we can compute the ``naive'' R-charge
assignments for the bifundamentals. Here we indicate the conjectured R-charges for the top cell of
Gr$^{\mathrm{tnn}}_{k,n + k}$. At sufficiently large $k$ and $n$, these values will be corrected due to the
appearance of accidental $U(1)$'s in the IR theory. To avoid clutter, we only indicate some of the R-charges;
the remaining ones follow from equations \eqref{faceconstraint} and
\eqref{vertexconstraint}. All of the R-charges are proportional to $x = 2/(n+k)$.}%
\label{higherR}%
\end{figure}

Following the procedure of \cite{Kutasov:1995np, Kutasov:2003iy,
Intriligator:2003mi}, we can postulate the existence of an extra $U(1)$ which
acts only on the ``offending'' operator so that it can decouple as a free field as
it hits the unitarity bound. Including a sequence of such extra $U(1)$'s (as each further meson
decouples), the form of the trial a-function is now:
\begin{equation}
a_{\text{new}}\left[  x\right]  =a_{\text{old}}\left[  x\right]  +\frac
{3N^{2}}{32}\left(  \frac{2(r-1)}{9}+\frac{2(r-1)}{9}\right)
-a_{\text{left}}^{(r)}\left[  x\right]  -a_{\text{right}}^{(r)}\left[
x\right] \ ,
\end{equation}
where the subtracted contribution from the mesons is:%
\begin{equation}
\begin{split}
a_{\text{left}}^{(r)}\left[  x\right]   &  =\frac{3N^{2}}{32}\underset
{j=2}{\overset{r}{\sum}}  f\left(  z_{1}+x_{1}+...+x_{j}\right)  \ , \\
a_{\text{right}}^{(r)}\left[  x\right]   &  =\frac{3N^{2}}{32}\underset
{j=1}{\overset{r-1}{\sum}} f\left(  y_{n}+x_{n}+...+x_{n-j}\right) \ .
\end{split}
\end{equation}
Extremizing $a_{\text{new}}\left[  x\right]  $, the R-charges now organize
according to exterior regions, for $0\leq i\leq r$ or $n-r+1\leq i\leq n$
and an interior region, for $r<i<n-r$. All of the R-charges for the interior $X_{i}$ are identical,
while the R-charges for the exterior $X_{i}$ obey a sequence of nested
square root functions. A full analysis of these solutions is beyond the scope of the
present paper. As should be clear, these theories exhibit a surprisingly rich
phase structure which deserves further study.

Similar considerations apply in the Gr$_{k,n+k}^{\text{tnn}}$ theories with
$k>2$. In figure \ref{higherR} we present a conjecture for the
\textquotedblleft naive\textquotedblright\ R-charge assignments of the theory.
As $k$ and $n$ grow, however, we can expect mesons to decouple, which will in
turn modify the computation of the infrared R-symmetry, leading to
a gradation of R-charge assignments as one passes from
the outside of the graph to the interior. Qualitatively this is similar
to a variational problem for the area of a two-dimensional membrane.

\section{Bipartite Graphs and $\mathcal{N}=2$ BPS Quivers \label{sec:N=2}}

Although our main interest in this paper is the $\mathcal{N}=1$ theory
defined by a given bipartite graph, much of the underlying geometric structure
borrows from the string construction of generalized Argyres-Douglas (AD)
$\mathcal{N}=2$ theories. In preparation for our stringy realization of the
$\mathcal{N}=1$ bipartite theories, in this section we show that much of this
structure is also present in $\mathcal{N}=2$ theories, and in particular the
BPS quiver associated with a given theory.

The BPS quiver is a convenient way to capture data about 1/2 BPS states of an underlying $\mathcal{N} = 2$ theory.
In this subsection we explain how these quivers are related to our bipartite graphs. The
main statement is that in an appropriate chamber, the BPS quiver of the
generalized AD\ theory of type $(A_{k-1},A_{n-1})$ realizes the
internal nodes of the bipartite theories associated to Gr$^{tnn}_{k,n+k}$.
Most of the observations we make in this section are well-known to experts and
can be found for example in \cite{Cecotti:2010fi, Alim:2011ae, Xie:2012dw, Gaiotto:2012rg}. See
also \cite{Gaiotto:2012db} for some discussion of the relation between bipartite graphs
and spectral networks. The main distinction
from the $\mathcal{N} = 2$ case is the absence of
flavor symmetries for the planar bipartite graphs. This will automatically appear
in our stringy realization of the $\mathcal{N} = 1$ theories.

The $\mathcal{N}=2$ generalized $(A_{k-1},A_{n-1})$ AD\ theories
can be obtained in a number of ways from string theory. They arise from
compactification of type IIB strings on the singular Calabi-Yau threefold $y^{k}%
=x^{n}+uv$ (see e.g. \cite{Shapere:1999xr, Cecotti:2010fi}), which in the mirror IIA
geometry is characterized by an NS5-brane wrapped on the singular algebraic curve:
\begin{equation}
f(x,y)=y^{k}-x^{n}%
\end{equation}
in $\mathbb{C}^{2}$, with holomorphic coordinates $x$ and $y$. In the M-theory lift this
corresponds to an M5-brane wrapped on the same curve. Throughout we refer to this algebraic
curve and its deformations as $\Sigma$.

The Coulomb branch of the $(A_{k-1},A_{n-1})$ AD\ theory corresponds in the
geometry to switching on specific lower order deformations of the form
$x^{i}y^{j}$ in the geometry. The charge lattice of this theory has dimension
$(k-1)(n-1)$.\footnote{The dimension
of the charge lattice is equal to $2n_r+n_f$, where $n_r$ is the
dimension of the $\scN=2$ Coulomb branch and $n_f$ is the number of
$\scN=2$ mass parameters.}
A 1/2 BPS\ state is specified by a choice of $\mathcal{N}=1$ subalgebra of the
$\mathcal{N}=2$ theory, corresponding (in the M5-brane realization of the theory)
to M2-branes wrapped over two-chains ending on homology 1-cycles of $\Sigma$. The
1/2 BPS states of this $\mathcal{N} = 2$ theory are encoded in BPS flow lines, and this data can in turn be packaged in
terms of a BPS quiver.

In the BPS flow line picture (see \cite{Shapere:1999xr}), one tracks
the constant phase trajectories of the Seiberg-Witten differential
$\lambda_{SW} = y dx$. This is locally specified by the flow equation:
\begin{equation}
y dx = \alpha dt
\end{equation}
for $\alpha = \exp(i \theta)$ a constant phase which specifies a choice of $\mathcal{N} = 1$ subalgebra. Such flows meet at the
ramification points of $\Sigma$, viewed as a $k$-sheeted cover over the $x$-plane. A homology 1-cycle which encircles two
such ramification points specifies a BPS state of the theory. This projection to the $x$-plane is equivalently characterized
by the spectral network of the theory \cite{Gaiotto:2012rg}.

The spectrum of 1/2 BPS states of the generalized AD\ theory is also encoded in the BPS quiver of the
$\mathcal{N}=2$ theory \cite{Cecotti:2010fi, Alim:2011ae}. This is just the
quiver diagram we would get from characterizing possible bound states of
branes in the singular Calabi-Yau $y^{k}=x^{n}+uv$. Depending on the details
of the deformations of the singularity, the connectivity of the quiver can be
different, corresponding to a mutation i.e. a Seiberg-like duality of
the quiver.

\begin{figure}[t!]
\centering{\includegraphics[width=0.5\linewidth]{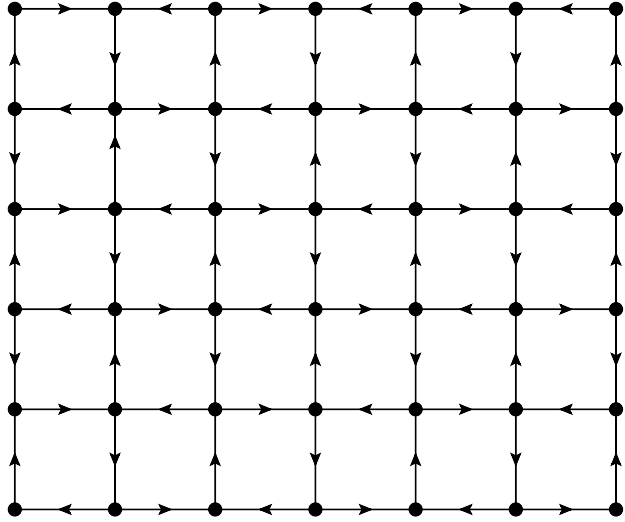}}
\caption{A BPS quiver for the $(A_{k-1}, A_{n-1})$ Argyres-Douglas theory. This is given by a $(k-1) \times (n-1)$ grid of nodes,
which are attached as in the figure. The rectangular grid of squares specifies the canonical chamber of the theory. Other
chambers can be reached by a sequence of mutations (see figure \ref{mutation}).}
\label{fig.BPSquiver}
\end{figure}

\begin{figure}[t!]
\centering{\includegraphics[scale=1.7]{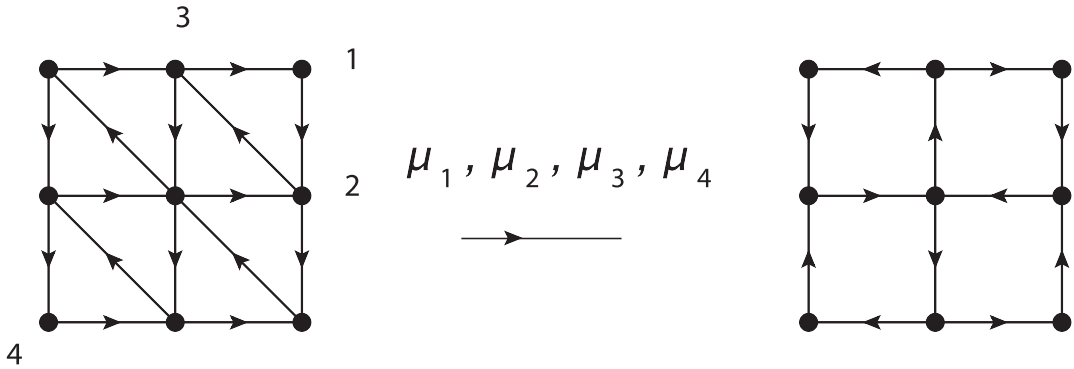}}
\caption{By a sequence of mutations, the internal nodes of the quiver from figure $\ref{higher}$ for the
top cell of the Gr$^{\mathrm{tnn}}_{k,n+k}$ theories can be brought to
the canonical chamber of the $\mathcal{N} = 2$ $(A_{k-1}, A_{n-1})$ BPS quiver. In the figure we
present this transformation for $k = 4$ and $n = 4$ by mutation on the nodes labelled $1,...,4$.}
\label{mutation}
\end{figure}

The canonical chamber for the BPS quiver of the $(A_{k-1},A_{n-1})$ theory is
a $(k-1)\times(n-1)$ rectangular grid of quiver nodes in which each interior node supports two horizontal
arrows both pointing into or out of the node, and two vertical arrows
anti-correlated with the orientation of the horizontal arrows (see figure \ref{fig.BPSquiver}).
By tuning the deformation parameters of $\Sigma$, all of the BPS states
can be made mutually supersymmetric. Then, the quiver literally describes the BPS states interpolating between
ramification points of $\Sigma$. Note also that the resulting quiver can be described as a bipartite
graph with a grid of square faces. Even so, a given mutation of the quiver may not
be characterized by a bipartite graph. One can also show that the interior pattern of tiled hexagons depicted in figure \ref{higher}
can indeed be reached by a sequence of such mutations, see figure \ref{mutation} for an example.
With a little bit of patience, one can see that this is true for general $k$ and $n$.

Higgsing in the BPS quiver corresponds to a partial resolution of
the singular geometry, which in the quiver amounts to switching on FI parameters.
This in turn triggers non-zero bifundamental VEVs which in the bipartite graph corresponds
to deleting the neighboring edge between faces (see e.g. \cite{Franco:2005rj} for the related
statement in $\mathcal{N} = 1$ theories). Performing
several such Higgsing operations in tandem with mutations, one can see that
other bipartite BPS quivers can also be reached.

\begin{figure}[t!]
\centering{\includegraphics[width=\linewidth]{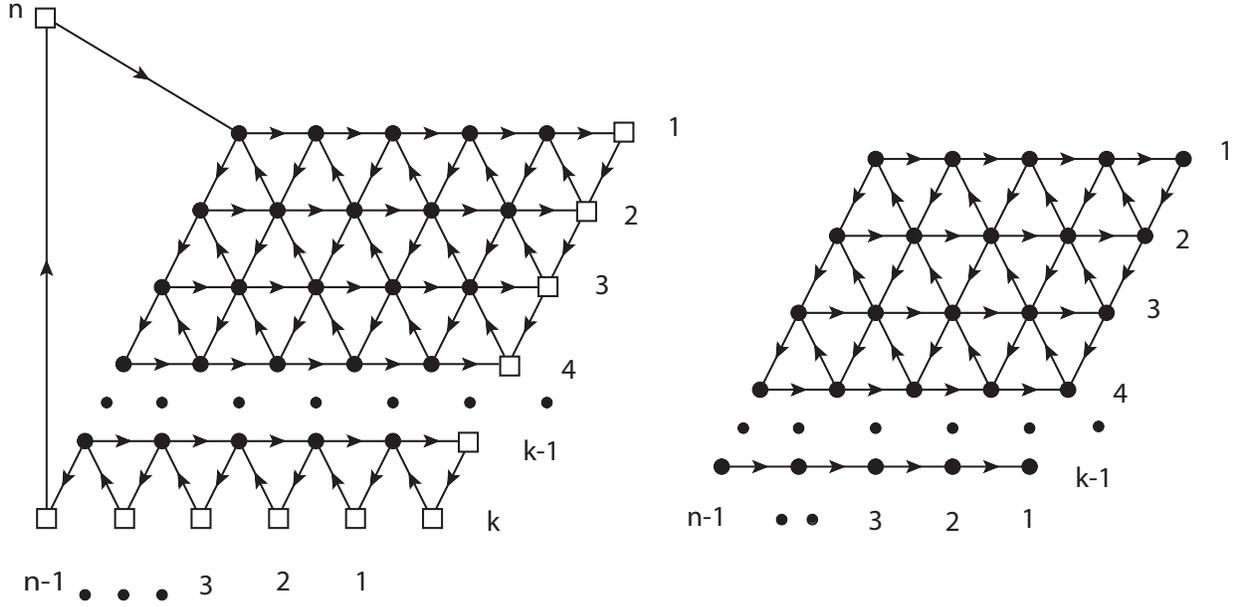}}
\caption{Comparison of the 4D $\scN=1$ quiver (left) and the BPS quiver for the
4D $\scN=2$ Argyres-Douglas theory (right). The main difference
between these two quivers is that the one for the $\scN=2$ theory
does not contain any flavor nodes, which we denote by squares.}
\label{highercompare}
\end{figure}

Summarizing, we see that the interior of a bipartite graph is indeed naturally reproduced from an $\mathcal{N} = 2$
BPS quiver. However, we can also identify some differences. If we had tried to interpret the BPS
quiver as a 4D gauge theory, there would not be any flavor nodes (see figure \ref{highercompare}) and
we would find that the gauge theory is either anomalous, or develops a mass gap. Hence,
a proper treatment of the flavor groups is essential in the definition of these theories. We
now turn to a stringy realization of the $\mathcal{N} = 1$ bipartite theories and show how the
flavor branes naturally appear.

\section{Stringy Realization \label{sec:STRING}}

In this section we propose a string theory realization for the
bipartite gauge theories introduced in section \ref{sec:REVIEW}.
The basic setup consists of a configuration of intersecting
D5-branes and NS5-branes wrapped on mutually supersymmetric
two-cycles in the internal geometry. Although the resulting
system only preserves $\mathcal{N}=1$ supersymmetry, it turns out that much of
the necessary geometric data can be formulated in terms of closely related
objects in a corresponding $\mathcal{N}=2$ AD theory. The basic idea will be to
consider a configuration of intersections between D5-branes and NS5-branes which
are characterized by BPS flow lines of an associated $\mathcal{N} = 2$ theory.

Our setup is as follows. We take type IIB\ string theory on $\mathbb{R}%
^{3,1}\times\mathbb{C}^{2}\times\mathbb{C}_{\bot}$, with D5-branes and
NS5-branes filling the four spacetime directions, and wrapping two-dimensional
subspaces in the internal geometry. All branes sit at the same point of
$\mathbb{C}_{\bot}$. With respect to this fixed complex structure for
$\mathbb{C}^{2}$, we take $N$ D5-branes wrapping an algebraic curve
$\Sigma\subset\mathbb{C}^{2}$ with specified boundary conditions at infinity
in $\mathbb{C}^{2}$. In addition to these D5-branes, we consider NS5-branes
wrapped on non-compact special Lagrangian submanifolds inside of
$\mathbb{C}^{2}$ which intersect the D5-branes along 1-cycles of $\Sigma$.
These intersections serve to partition the D5-brane into independent pieces.
Indeed, once the NS5-branes have been added, the D5-branes can be viewed as
separate branes which are suspended between the NS5-branes. In this configuration,
there are moduli associated with the VEVs of massless bifundamentals
stretched between the various D5-branes. These directions in the moduli space
correspond to the motion of the distinct suspended D5-branes in the configuration.

The algebraic curve $\Sigma$ wrapped by the D5-branes is the zero set $f(x,y) = 0$ with:%
\begin{equation}
f(x,y)=y^{k}-x^{n}+\sum a_{ij}x^{i}y^{j}\label{fxy} \ ,%
\end{equation}
where $x$ and $y$ are holomorphic coordinates of $\mathbb{C}^{2}$ and the
terms multiplying $a_{ij}$ are lower degree deformations. Note that in the absence
of the NS5-branes, the D5-branes would preserve $\mathcal{N}=2$ supersymmetry. In
contrast to an M5-brane wrapped over the geometry with all deformations
switched off, the resulting IR theory produced by the D5-brane would
not describe an interacting fixed point. The reason
is that Euclidean D1-branes wrapped on two-chains ending on one-cycles
will in general smooth out the singular behavior of the classical geometry (see e.g.
\cite{Ooguri:1996me}). Compactifying on a further circle, however, leads to a
three-dimensional theory which is identical to the circle compactification of
the $\mathcal{N}=2$ $(A_{k-1} , A_{n-1})$ AD theory.

Adding NS5-branes wrapped on mutually supersymmetric special Lagrangians
breaks the supersymmetry further, leaving us with a 4D $\mathcal{N}=1$ theory.
This choice of an $\mathcal{N}=1$ subalgebra has a close cousin in the related
$\mathcal{N}=2$ theories: It specifies BPS flow lines of the generalized AD theory.
In the case at hand, a network of BPS flows determines a set of special Lagrangian submanifolds in $\mathbb{C}^2$
which are wrapped by NS5-branes. Recall that a special Lagrangian is specified by the
conditions:%
\begin{equation}
J|_{L}=0\text{ and }\operatorname{Im}e^{-i\theta}\Omega|_{L}=0 \ ,
\end{equation}
where $J = \frac{i}{2}(dx \wedge d \overline{x} + dy \wedge d \overline{y})$
denotes the K\"{a}hler form of $\mathbb{C}^{2}$ and $\Omega=dx\wedge
dy$ is the holomorphic two-form. The parameter $\theta$ specifies a choice of
$\mathcal{N}=1$ subalgebra which is preserved. Integrating the holomorphic two-form
over a contour normal to $\Sigma$ yields the Seiberg-Witten
differential $\lambda_{SW} = y dx$. Given such a phase $\alpha=\exp(i\theta)$,
there is a corresponding BPS flow line satisfying the differential equation:
\begin{equation}
ydx = \alpha dt \  .
\end{equation}
Conversely, given such a BPS\ flow line, there exists a special Lagrangian in
$\mathbb{C}^{2}$ which intersects $\Sigma$ along this flow line. See Appendix
B for how to reconstruct a special Lagrangian in $\mathbb{C}^2$ given a BPS
flow line on $\Sigma$.

Partitioning the D5-brane up into distinct components realizes an
$\mathcal{N} = 1 $ quiver gauge theory. Each disconnected component
on a sheet of $\Sigma$ specifies either a flavor group, or a gauge
group. More precisely, to maintain dynamical gauge groups we
assume that at large $x$ and $y$ the geometry has been cut off.
Dynamical gauge group factors correspond to at most logarithmic running with
the cutoff, while flavor groups exhibit power law running, a feature we explain further in subsection \ref{ssec:UV}.

In addition to these symmetry groups, when two disconnected regions share a one-dimensional
interface, there will be a bifundamental chiral superfield localized there.
When three or more such 1-cycles of the network meet at a point, a
worldsheet disk instanton localized at the intersection point generates a
superpotential interaction term between the corresponding bifundamentals.
This gives rise to a quiver gauge theory, with superpotential interaction
terms dictated by the appearance of more than two
lines meeting.

As should now be clear, our construction is closely related to the BPS
quiver of the $\mathcal{N} = 2$ AD theories. Indeed, while the interiors of the two quivers
are the same, there is an important distinction in the appearance of flavor branes for
the $\mathcal{N} = 1$ theories.  We now explain why the interior of the $\mathcal{N} = 1$
quiver gives the BPS quiver for $\mathcal{N} = 2$ theories.  As we
change the phase of the BPS flow lines, the topology of the flow lines
change precisely at phases corresponding to where they are
given by the phase of the central charge of the $\mathcal{N} = 2$
BPS state.  Consider the case where we take the parameters of the
$\mathcal{N} = 2$ curve to give almost the same phase for all the BPS states.
In this limit the interior regions collapse to regions with zero area.
In this limit the theory one gets for the $\mathcal{N} = 1$ quiver,
if we T-dualize three of the spatial dimensions and keep only
the interior branes is the
$\mathcal{N} = 2$ BPS quiver\footnote{Note that this also explains why
the $\mathcal{N} = 1$ theory would have been anomalous if
we had just kept the interior regions of the quiver, since T-dualizing all spatial
directions would yield a situation where the flux due to BPS charges
has nowhere to go.}.  This thus explains why the interior part of the
${\cal N}=1$ quiver agrees with that of the ${\cal N}=2$ BPS quiver. The geometry of the
flavor branes is fixed by the asymptotics of the BPS flow lines at large $x$ (c.f. Appendix B),
which partitions a sheet of the cover into $n+k$ distinct asymptotic regions. In other words, combining the asymptotic behavior of the
flows with the interior behavior predicted by the BPS quivers, we reconstruct the general
pattern of quiver gauge group factors found in the bipartite theories! More precisely, we recover the
most generic class of quiver gauge theories, corresponding to the top
cell of Gr$^{\rm tnn}_{k,n+k}$.

Reaching the lower cells of Gr$^{\rm tnn}_{k,n+k}$ is also straightforward in this
construction. In the bipartite graph, we reach these cells by deleting some
of the internal lines of the bipartite graph. In the quiver gauge theory,
deleting a line corresponds to activating a pair of FI\ parameters on the two
faces which share this edge \cite{Franco:2005rj}. These FI parameters
are given by background values of two-form potentials integrated over
appropriate two-chains. Setting these FI\ parameters to $\zeta$ and
$-\zeta$ triggers a VEV for the bifundamental which is consistent (via
earlier results on dimer models) with all F- and D-term constraints.
These VEVs then break the two gauge groups to its diagonal via the Higgs
mechanism.

In the remainder of this section we discuss in more detail the construction of
these quiver gauge theories. To give a set of illustrative examples, in
subsection \ref{ssec:EXAMPLE} we show in detail how our construction
reproduces the cells of Gr$^{\rm tnn}_{2,n+2}$. This case illustrates the main
aspects of the general construction, and also reveals that the white and black
vertices attached to external legs are indeed treated differently in our
construction. In subsection \ref{ssec:GENERALIZE} we exploit the geometric connection with related
$\mathcal{N}=2$ theories to show how the Gr$^{\rm tnn}_{k,n+k}$ theories are
reproduced from our construction. This is followed by a more detailed discussion of the
cutoff geometry, and the class of admissible deformations of $\Sigma$ in subsection \ref{ssec:UV}.
Finally, making use of this geometric
characterization of our theories, in subsection \ref{ssec:SEIBERG} we show that
the local combinatorics such as the \textquotedblleft square
move\textquotedblright\ (i.e. Seiberg duality) is reproduced by the brane construction. This establishes that the IR
fixed points are indeed characterized by the cells of the Grassmannian.

\subsection{The Gr$^{\rm tnn}_{2,n+2}$ Theories \label{ssec:EXAMPLE}}

\begin{figure}[t!]
\centering{\includegraphics[scale=1.2]{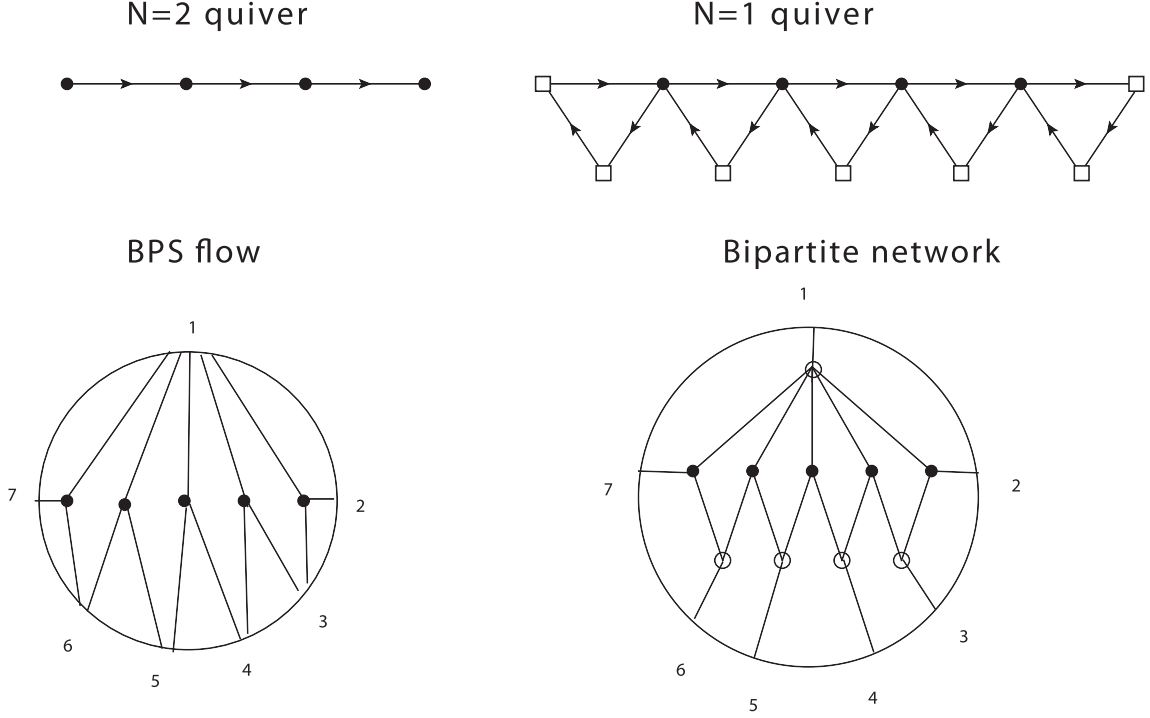}}
\caption{Depiction of the $(A_{1} , A_{n-1})$ $\mathcal{N} = 2$ BPS quiver, the corresponding $\mathcal{N} = 1$
quiver, and the flow lines for theories associated with the top cell of Gr$^{\mathrm{tnn}}_{2,n+2}$.
Additionally, we present the corresponding bipartite network. For specificity, in the figure we indicate the case
of $n = 5$. See figure \ref{canonical} for the R-charge assignments of the $\mathcal{N} = 1$ theory, and figure \ref{floweg} for
another example of flow lines and the corresponding bipartite network.}
\label{fig.G2flow}
\end{figure}

In this subsection we show how our proposal works in more detail in the
special case $k=2$ so that the curve $\Sigma$ is given by:%
\begin{equation}
y^{2}=P_{n}(x)=\underset{i=1}{\overset{n}
{{\displaystyle\prod}
}}(x-a_{i}) \ .
\end{equation}
This is the curve for the $(A_{1} , A_{n-1})$ Argyres-Douglas theories. The BPS quiver for these theories consists of a
single row of $(n-1)$ nodes. What we are going to show is how our $\mathcal{N} = 1$ construction based on BPS flow lines
naturally supplements this quiver by additional flow lines. In particular, this will convert the quiver nodes which would have been anomalous
in a 4D theory into anomaly free gauge group factors. See figure \ref{fig.G2flow} for
a depiction of the $\mathcal{N} = 2$ BPS quiver, the $\mathcal{N} = 1$ quiver,
and the corresponding flow lines and bipartite graph.

Recall that the BPS\ flow lines of $\Sigma$ are given by constant phase trajectories of
$\lambda_{SW}=ydx$. Introducing a local parametrization $t$ for the flow,
this can locally be written as the condition:%
\begin{equation}
ydx=\alpha dt \label{flowdiff}%
\end{equation}
for $\alpha=\exp(i\theta)$ a choice of phase for the sLag. This constant phase
condition has been studied in the context of BPS flow lines of $\mathcal{N}=2$
systems \cite{Shapere:1999xr}, and also appears in the WKB\ approximation of
the complexified quantum mechanics with Hamiltonian $H(x,p)=p^{2}-P_{n}%
(x)$.\footnote{\label{footnote:QM}In the complexified quantum mechanics, the BPS flow lines
specify saddle point configurations of the complexified path integral:%
\begin{equation}
\int\mathcal{D}\left[  x(t)\right]  \mathcal{D}\left[  p(t)\right]
\exp\left(  i\int pdx-H(x,p)dt\right)
\end{equation}
where here, $H(x,p)$ is a general holomorphic polynomial in $x$ and $p$. The
complexification is similar in spirit to the discussion in \cite{Witten:2010zr}.
A level set $H= E$ defines a Riemann surface in the $\mathbb{C}^{2}$
parameterized by $x$ and $p$. A stationary phase of the path
integral specifies a complex trajectory in the $(x,p)$ phase space
which we can denote by $(x(t),p(t))$. In this system, we observe that along a level set,
the stationary phase configuration corresponds to the
case where $\int pdx$ has a fixed phase along a given trajectory. In other
words, the BPS\ flow line specifies the WKB\ approximation of the complexified
quantum mechanics.}

The asymptotic geometry of the BPS\ flow lines at large $x$ and $y$ is
dominated by the highest degree term $x^{n}$ of $P_{n}(x)$. Integrating the
flow equation in this case yields the value of $x$, treated as a function of
$t$:
\begin{equation}
x^{(n+2)/2}-x_{0}^{(n+2)/2}=\beta t \ ,
\end{equation}
where $\beta$ is a constant with the same phase as $\alpha$, and $x_{0}$ is
the value of $x$ at $t=0$. In other words, the asymptotics of the one-cycle
are fully specified by the phase $\exp(i\theta)$. We see that since
$t\in \mathbb{R}$, there are precisely $n+2$ asymptotic regions for the one-cycles, given by
$x\sim r\zeta_{n+2}$ for $r\gg0$ and $\zeta_{n+2}$ an $(n+2)$-th root of
unity. This specifies a boundary condition for the NS5-brane, which remains
fixed once we specify a choice of $\mathcal{N}=1$ algebra.

In the interior of the $x$-plane these asymptotic flows
will meet at roots of $P_{n}(x)$. Near a given root
of $P_{n}(x)$, we can locally integrate equation (\ref{flowdiff}).
If the root of $P_{n}(x)$ vanishes to order $m$, this yields
the local behavior of the flow near this root:%
\begin{equation}
(x-a_{i})^{(m+2)/2}\simeq\beta^{\prime}t^{\prime}%
\end{equation}
where $\beta^{\prime}$ is a constant with the same phase as $\alpha$.
This locally defines the meeting of $m+2$ BPS\ flow lines. In particular, we
see that in the generic case where $m=1$, this specifies a trivalent vertex.

These trivalent vertices naturally map to vertices of a bipartite graph.
Indeed, each region of the $x$-plane divided by the flow lines specifies a
gauge group factor, and D5/D5$^{\prime}$ strings stretched across the
interface provide matter fields. Note that since the BPS\ flow lines come with
a choice of local orientation, the GSO\ projection leads to a chiral
superfield with a designated orientation in the quiver gauge theory.
Superpotential interactions generated by a worldsheet disk instanton localize
at the root $x=a_{i}$. In other words, this reproduces one of the building
blocks of the bipartite theories.

But in addition to regions where the BPS\ flow lines meet at a root, there can
also be regions where the flow lines \textit{nearly} meet. In this situation,
two or more BPS\ flow lines in the interior asymptote to the same root of
unity at large $x\sim r\zeta_{n+2}$. Tracing these flows back into the
interior, one can see that although they come close to touching, they
eventually diverge from one another. Asymptotically, one can still view this
as a trivalent vertex, but one in which neighboring non-compact faces do
\textit{not} contribute an additional bifundamental. Consequently, there is
also no superpotential interaction term associated with the corresponding
vertex. In other words, we see that the BPS flow lines either terminate at
trivalent vertices, or form asymptotic flows.

\begin{figure}[t!]
\centering{\includegraphics[scale=0.35]{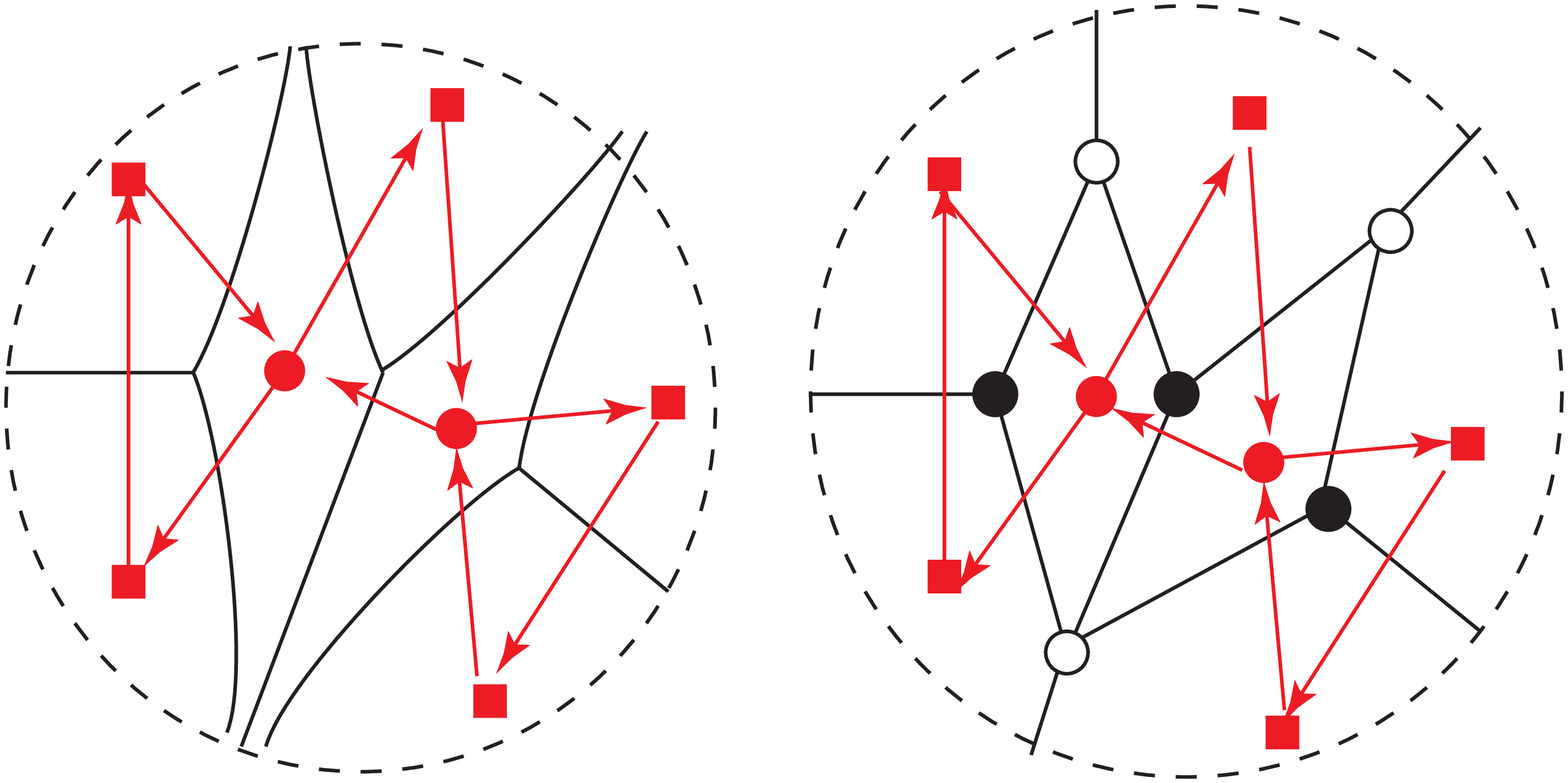}}
\caption{Depiction of the flow lines and corresponding bipartite graph realized by a configuration of
D5-branes wrapping $y^2 = x^3 + (\mathrm{deformations})$, with NS5-branes intersecting the D5-brane
along the BPS flow lines of the related $\mathcal{N} = 2$ $(A_{1} , A_{2})$ Argyres-Douglas theory.}
\label{floweg}
\end{figure}

These flows come in two types, and for the class of $y^{2}=P_{n}(x)$ theories
reproduce the bipartite graph associated with the top cell
of Gr$^{\rm tnn}_{2,n+2}$. To see this, order the roots of $P_{n}(x)$ so that
$\operatorname{Re}a_{1}\leq\operatorname{Re}a_{2} \leq... \leq
\operatorname{Re}a_{n-1}\leq\operatorname{Re}a_{n}$. The trivalent vertices
produce a pattern of diamonds which are glued pairwise along a
common edge. See figure \ref{floweg} for a depiction of the flow lines for
the case $k=2$ and $n=3$. The generalization to higher $n$ for a particular
choice of duality frame is shown in figure \ref{fig.G2flow}. The pattern of flow lines
maps out a collection of ``diamonds'' which are glued together to form a single row of squares.
Let us note that in general, there is no unique bipartite graph constructed from these flows. For
example, by moving the roots along the imaginary axis, we can either produce a row of diamonds which
all share a single white vertex, or alternatively, we can produce a row of squares which only share adjacent
sides. In any case, the external data is the same: we obtain a bipartite graph with an
external leg attached to the leftmost black vertex and another external leg
attached to the rightmost black vertex, with the $n$ remaining legs attached to
white vertices. In other words, we have found the bipartite graphs associated
with the top cell of Gr$^{\rm tnn}_{2,n+2}$! As already mentioned, we arrive
at the lower cells by switching on background values for field-dependent FI parameters.
This triggers a brane recombination, corresponding to deleting internal lines of the bipartite graph.

An important feature of our construction is that the flavor group factors are
not anomalous which means we are free to weakly gauge these symmetries. This
is because the \textquotedblleft external lines\textquotedblright\ associated
with the white vertices do not contribute a bifundamental mode. Thus, we
conclude that our construction realizes the rules for external legs proposed in
\cite{Xie:2012mr}.

\subsection{The Gr$^{\rm tnn}_{k,n+k}$ Theories \label{ssec:GENERALIZE}}

In this subsection we show how our construction
generalizes to the Gr$^{\rm tnn}_{k,n+k}$ theories,
corresponding to taking the curve $\Sigma$
associated with the $(A_{k-1},A_{n-1})$ AD theories. In this case, we can view
$\Sigma$ as a $k$-sheeted cover over the $x$-plane, with ramification points
at the locations where different sheets meet. Throughout, we work in terms of
the projection down to the $x$-plane.

The trajectory of the BPS\ flow lines is again dictated
by the condition that $\lambda_{SW}=ydx$ has a constant phase.
One important difference from the $k=2$ case is that
$y$ has $k$ branches $y_1(x), y_2(x), \ldots, y_{k}(x)$. Correspondingly
we could consider $k(k-1)/2$ Seiberg-Witten differentials
$\lambda_{ij}:=y_i dx-y_jdx$, and the same number of flow lines labeled
by $(ij)$:
\beq
\lambda_{ij}=(y_i-y_j) dx =e^{i\theta} dt \ ,
\eeq
with constant $\theta$. Note that the flow lines
come with a natural orientation, such that the
value of $\textrm{Re}\,(e^{-i\theta} \lambda_{ij})$ increases along a
flow. Obviously the orientation flips when we exchange $i$ with $j$.

Let us for the moment concentrate on one of the flow lines.
Away from the ramification points, we can see that the asymptotics of the BPS flow line
picture will be quite similar to the case $k=2$. Integrating the flow equation
yields the value of $x$, treated as a function of $t$:
\begin{equation}
x^{(n+k)/k}-x_{0}^{(n+k)/k}=\beta t \ ,
\end{equation}
so that now there are $n + k$ asymptotic regions on the sheet. Continuing this
asymptotic behavior into the interior of the $x$-plane, we can see that some
of these flow lines will remain as a single line, while others will split
apart, corresponding to lines which nearly touch. Just as in the $k=2$ case,
these two possibilities correspond to external lines which either attach to a
black vertex or a white vertex. This in turn specifies the details of possible
bifundamentals connecting the flavor branes.

\begin{figure}[t!]
\centering{\includegraphics[scale=1.2]{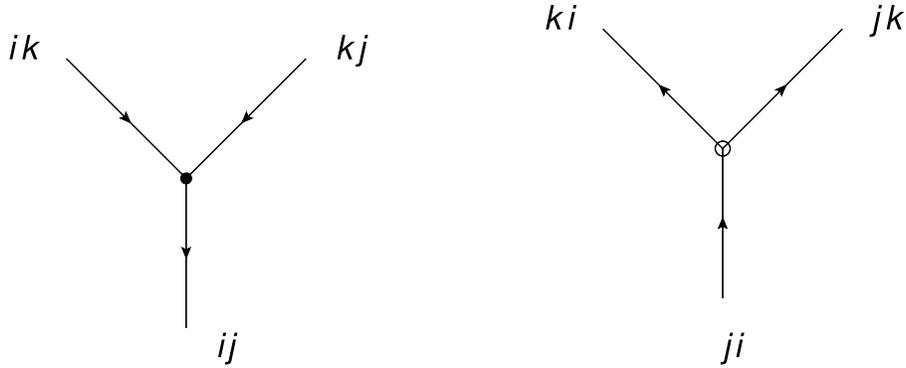}}
\caption{Two kinds of trivalent vertices for the BPS flow lines at the
 branch points, with different orientations.}
\label{fig.vertex}
\end{figure}

Going into the interior region, it should be clear that the global structure of the flows
will be more involved than in the $k = 2$ case. The main physical content we need to track corresponds
to the meeting of flow lines near a ramification point of the curve $\Sigma$. This interior
structure is in turn dictated by the spectral network of the related $\mathcal{N} = 2$
theory \cite{Gaiotto:2012rg}. Before turning to this global characterization, let us
first discuss the local structure of flows which meet. The generic situation corresponds
to the meeting of three BPS flow lines at a ramification point. Near this meeting
point, the sheets which appear in the joining operation satisfy the relation:
\begin{equation}
\lambda_{ij}+\lambda_{jk}=\lambda_{ik}.
\end{equation}
Hence, we see that the triple intersection point satisfies the join and split rule
for strings: $(ij)+(jk)\rightarrow (ik)$. Moreover, there are two possible types of
orientations for these flow lines, which we show in figure \ref{fig.vertex}. The
reason is that if $(ij)$ and $(jk)$ come into the triple intersection point, then
the orientation of $(ik)$ is fixed as outgoing. Alternatively, if $(ij)$ and $(jk)$
are outgoing, then the orientation of $(ik)$ is fixed as incoming. Note that this
is the \textit{same} choice of orientations as in figure \ref{fig.sinksource}(a)
in Appendix A. This means that the perfect orientation discussed in
Appendix A is nothing but the orientation of the flow lines!

\begin{figure}[h!]
\centering{\includegraphics[scale=1.6]{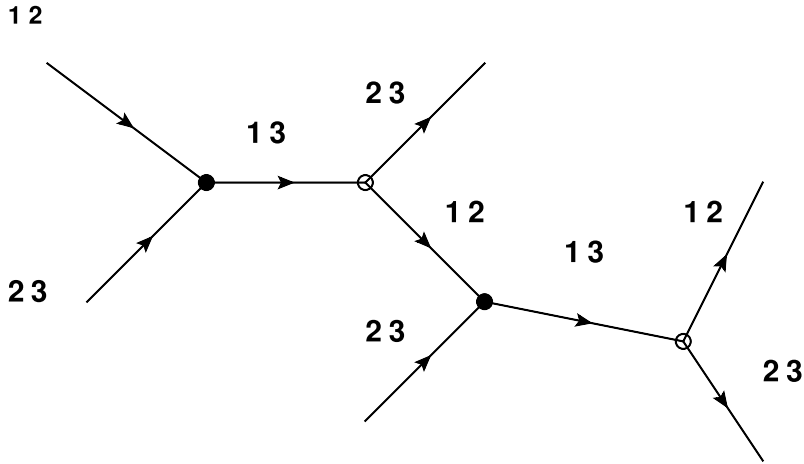}}
\caption{The flow skeleton for the $k=3$ theory.}
\label{orientation}
\end{figure}

Gluing this local data together, we obtain a network of flow lines which
meet at the triple intersection points. For this purpose we can combine the
trivalent vertices in figure \ref{fig.vertex} to
construct a ``skeleton'' of the flow lines, as in figure
\ref{orientation}. This generates a network which is equipped
with a perfect orientation.

\begin{figure}[t!]
\centering{\includegraphics[scale=1.4]{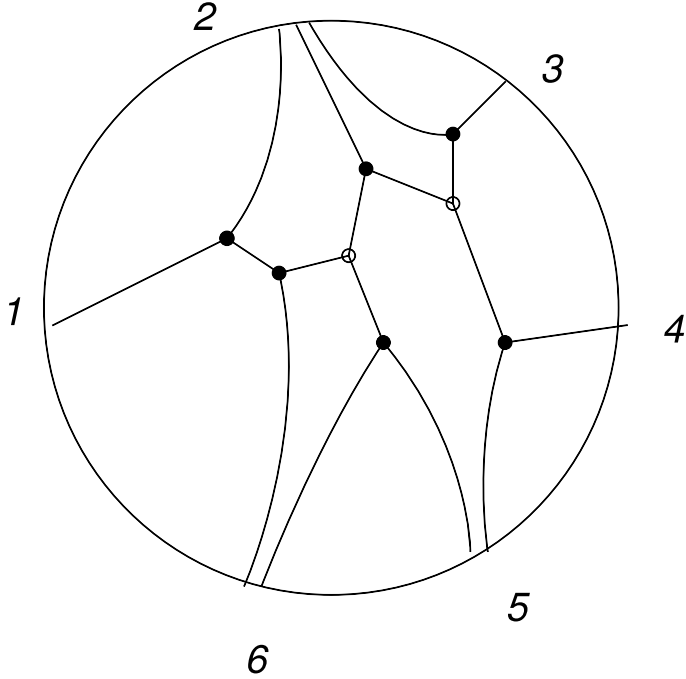}}
\caption{An example of the global structure of BPS flow lines, with $k=3, n=3$.}
\label{higherflow}
\end{figure}

Let us now come back to the global structure of the flow lines, such as
the one given in figure \ref{higherflow}.  We need to
extract the internal connectivity of these theories, and in
particular the adjacency of the corresponding quiver gauge theory realized by
the network of flow lines. Specifying how all of the sheets
glue together more globally is clearly more challenging, as it involves the
construction of the spectral network for the general $(A_{k-1} , A_{n-1})$
Argyres-Douglas theory. We leave a direct construction
of this type for future work. Instead, we follow a different
route to extract the corresponding quiver gauge theory specified by the flow lines.

Rather than perform a direct construction, we can
combine our knowledge of the asymptotic behavior of the BPS flows with the known internal structure of the
quiver to extract the corresponding gauge theories. Indeed, what
makes the analysis tractable is the fact that the BPS\ quivers for the
$(A_{k-1},A_{n-1})$ theories are already known \cite{Cecotti:2010fi}. In other words, we already know
the \textit{internal} structure of the quiver gauge theory we are going to get. Further, since we also know that the asymptotic flows terminate on
ramification points of the curve $\Sigma$, we can uniquely extend a bipartite chamber of the $\mathcal{N} = 2$ BPS quiver
to also include flavor branes. Note that this is in accord with the construction we have given for the
Gr$^{\rm tnn}_{2,n+2}$ theories.

Putting this together, we see that the construction just given realizes
the bipartite theories defined by the top cell of Gr$^{\rm tnn}_{k,n+k}$.
The bipartite graph will also coincide with the skeleton of the
flow lines introduced previously. Moreover, as we have already mentioned, we reach the lower
cells by a partial Higgsing of the quiver gauge theory. Thus, we see
that our construction recovers all of the possible bipartite theories!

Using the characterization of the interior using the BPS\ quiver, we can also
see that the Grassmannian duality $k\leftrightarrow n$ retains the same
connectivity of the bipartite graph, but with the black and white vertices
interchanged. The asymptotic flow lines, however, will be different. Indeed,
though the number of flavor branes remains the same, we can also see that the
bifundamentals connecting them will be distinct. This is in accord with the
fact that in the bipartite theories of \cite{Xie:2012mr}, the theories for
Gr$^{\rm tnn}_{k,n+k}$ and Gr$^{\rm tnn}_{n,n+k}$ correspond to different fixed points.

\subsection{UV\ Boundary Conditions \label{ssec:UV}}

So far we have shown that the bipartite graphs associated with Gr$^{\rm tnn}_{k,n+k}$
are naturally reproduced by the configuration of BPS flow lines for the
$(A_{k-1},A_{n-1})$ AD\ theories. From the structure of the BPS flow lines we
can also see, however, that the region enclosed by a \textquotedblleft
compact\textquotedblright\ face of the bipartite graph can actually be
non-compact (see e.g. figures \ref{fig.G2flow} and \ref{floweg}). Further, since the normalization of the D-term
$Q^{\dag}Q$ for a bifundamental $Q$ is proportional to
the length of the corresponding edge, we also see that
bifundamentals which are localized along flow lines may also be non-dynamical.

To realize a dynamical quiver gauge theory, it is therefore necessary to cut
off our non-compact geometry which in turn specifies a choice of UV\ boundary
conditions for the brane configuration. The basic idea is to evaluate the area
enclosed by a region partitioned by the BPS\ flow lines, using the restriction
of the K\"{a}hler form:%
\begin{equation}
J=\frac{i}{2}\left(  dx\wedge d\overline{x}+dy\wedge d\overline{y}\right)
\text{.}%
\end{equation}
In the large $x$ regime on the $x$-plane, the contribution from $dy\wedge
d\overline{y}$ is subleading, so it is enough to compute the area cut out in
the $x$-plane by the BPS\ flow line.

We now illustrate how to estimate this area and
the dependence on a choice of cutoff in the case
of the square dimer, specified by the curve $y^{2}=x^{2}-a^{2}$
for real $a>0$. In the large $x$ limit, the asymptotic behavior of $y$ is:%
\begin{equation}
y=\sqrt{x^{2}-a^{2}}\simeq x-\frac{a^{2}}{2x} \ .%
\end{equation}
Integrating this, the constant phase trajectories satisfy:%
\begin{equation}
x^{2}-a^{2}\log x\simeq\beta t
\end{equation}
for some constant $\beta$ with the same phase as $\alpha$, and flow coordinate $t$. The asymptotic
regime $x_{2}\rightarrow\infty$ and $x_{1}\rightarrow0$ for $x=x_{1}+ix_{2}$
imposes the condition:%
\begin{equation}
x_{2}\simeq\frac{4}{\pi a^{2}}\frac{1}{x_{1}} \ .
\end{equation}
In other words, the area underneath the curve $x_{2}(x_{1})$, viewed as a
function of $x_{1}$ depends logarithmically on the cutoff:%
\begin{equation}
\int_{\varepsilon}x_{2}(x_{1})dx_{1}\simeq-\frac{4}{\pi
 a^{2}}\log\varepsilon \ .
\end{equation}
This logarithmic running reflects the dependence of $1/g_{\rm YM}^{2}\sim \textrm{Area}(\textrm{Face})$ on the cutoff.

Similar considerations apply for the $(A_{k-1},A_{n-1})$ theories and their
deformations:%
\begin{equation}
y^{k}-x^{n}+\sum a_{ij}x^{i}y^{j}=0 \ .%
\end{equation}
In this case, the admissible deformations are those which only alter the UV
boundary conditions by at most logarithmically divergent terms. For example in the case
$k = 2$, deformations which induce at most logarithmic divergences in the area correspond to
degree $(n-2) / 2$ or lower terms in $x$. More generally, these lower order
deformations correspond in the $\mathcal{N} = 2$ theory to all deformations on the Coulomb branch,
and the lowest dimension mass deformation.

Including this cutoff dependence, it follows that the compact faces of the
bipartite graph correspond to dynamical gauge group factors in the 4D theory. By
contrast, we also see that the area of the flavor branes always diverges as a
power law in the cutoff. In particular, this means that we indeed have
realized the UV definitions of these quiver gauge theories.

\begin{figure}[t!]
\centering{\includegraphics[scale=0.5]{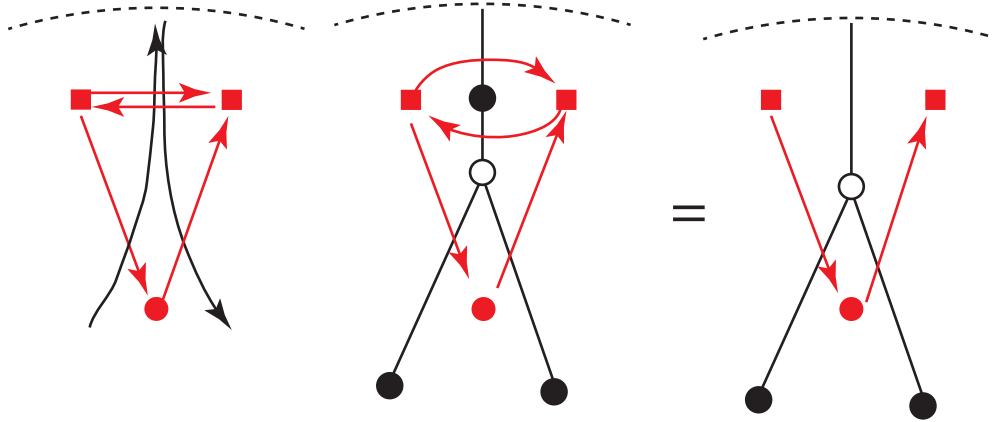}}
\caption{Cutting off the asymptotic behavior of the flows specifies UV boundary conditions
for the field theory. This in turn makes the gauge groups and some of the bifundamentals dynamical. Graphically,
this cutting off procedure can be viewed as attaching a black node to each external leg which attaches to a white node.
This gives a massive vector-like pair of bifundamentals, which in our construction is automatically integrated out.}
\label{fig.matterintegrate}
\end{figure}

It is also of interest to study the bifundamentals connecting flavor branes.
Once we introduce the cutoff, this can be viewed in the bipartite graph as
attaching an additional white node along each external leg which attaches to a
black vertex, and an additional black node for each external leg which
attaches to a white vertex. In particular, this shows that the asymptotic
white vertices can be viewed as also having an additional vector-like pair of
states, that is, massive string modes connecting the two adjacent flavor
branes. See figure \ref{fig.matterintegrate} for a depiction of this operation.

\subsection{Reproducing the Square Move \label{ssec:SEIBERG}}

Having shown that the bipartite gauge theories are correctly reproduced by our
construction, we can now ask whether the various combinatorial moves used to
classify possible IR\ fixed points are also reproduced. Most of the
combinatorial equivalences between different bipartite graphs follow from
statements in the corresponding weakly coupled gauge theory. For example, in a
bipartite graph with a node attached to only two lines, there is a
corresponding mass term available. These modes form a vector-like pair, and
are generically massive in the brane construction. In other words, they are
automatically integrated out.

The main case which is not apparent from the UV\ description of the weakly
coupled field theory is the \textquotedblleft square move\textquotedblright,
corresponding to the interchange of the black and white vertices, depicted in
figure \ref{fig.moves}(d). In the quiver gauge theory, this corresponds to a Seiberg
duality operation. We now show how to reproduce the square move from the BPS
flow line picture.

Much as in similar constructions, Seiberg duality of the field theory
corresponds in the brane configuration to a change of basis for the branes
which realize a given quiver gauge theory. In other words, holding fixed the
asymptotics of the intersecting branes, we consider lower order deformations
of the geometry. Since the net brane charge is fixed at infinity, the
infrared physics is the same. The realization of the quiver gauge theory could, however,
be different.

\begin{figure}[t!]
\centering{\includegraphics[scale=0.5]{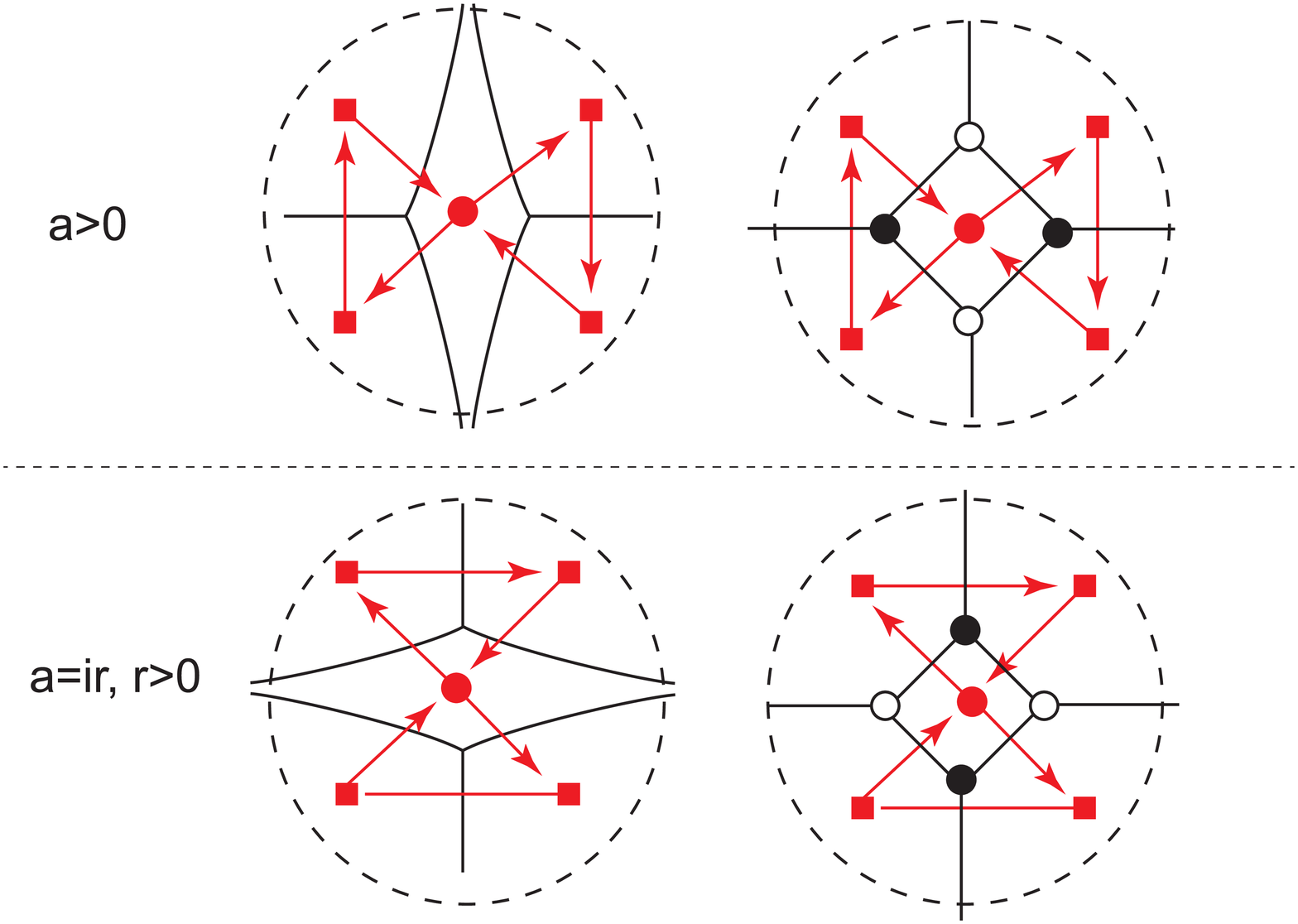}}
\caption{Seiberg duality in the brane construction corresponds
to a lower order deformation of the algebraic curve $\Sigma$. Here we depict
this operation for the curve $y^2 = x^2 - a^2$ as a function of the parameter $a$.
When we change $a$ from real $a>0$ to imaginary $a=ir, r>0$, the flow
lines of the curve $y^2=x^2-a^2$ are rotated by 90 degrees, which is
nothing but the square move of the bipartite network.}
\label{Seibergflow}
\end{figure}

The duality is achieved by varying the location of ramification points on the $x$-plane.
Under a generic variation, the geometry of the D5-brane remains smooth, but
the geometry of the NS5-branes can jump. As we show in Appendix B, given a BPS\ flow line, there is a
corresponding sLag which we can construct. So in other words, tracking the
change of the BPS flow lines is enough to study the possible jumping behavior
of the sLags.\ In the BPS\ quiver, this is simply a mutation of the quiver,
corresponding to moving to a different chamber of BPS states.

Since this jumping behavior of the flow lines is a local phenomenon, it is
enough to consider the case $y^{2}=x^{2}-a^{2}$, corresponding to the top cell
of Gr$^{\rm tnn}_{2,4}$, the single square dimer. When $a>0$, the six BPS\ flow
lines consist of two segments along the real $x$-axis, emanating from $x=a$ to
$x\rightarrow\infty$ and $x=-a$ to $x\rightarrow-\infty$. In addition, there
are four segments meeting these roots, which emanate out in pairs to either
$x\rightarrow i\infty$ or $x\rightarrow-i\infty$. As we rotate $a$ through the
complex plane, the asymptotics remain unchanged, but the internal connectivity
of the BPS flow lines will change. For example, when $a=ir$ with $r>0$, the
asymptotics remain the same, but internally, the entire flow picture has
rotated by $90$ degrees. Applying our previous rules for extracting a quiver
gauge theory, this corresponds to interchanging the black and white vertices
of the bipartite graph. See figure \ref{Seibergflow} for a depiction of Seiberg duality
in the geometry.

In the process of such deformations, one can also generically expect a brane
creation phenomenon. Note, however, that since the D5-branes are
\textquotedblleft passive\textquotedblright, i.e., they do not experience any
jumping, it should not be surprising that the rank of the gauge groups in the
Seiberg dual has not changed. Indeed, anomaly cancellation of the quiver gauge
theory and its Seiberg dual corresponds in the brane configuration to the
cancellation of RR and NS tadpoles. Let us note that in the more general
possibly non-conformal case where we allow different ranks on the various
gauge group factors, there can be a shift in the ranks of a given gauge group factor.

Let us summarize our discussion. We have seen that all of the local
combinatorial rules of the bipartite quiver gauge theories are correctly
reproduced by our string construction. Hence, the string construction provides a simple explanation for
\textit{why} different quiver gauge theories flow to the same interacting
fixed point.

\section{Physical Origin of Coordinates for Gr$^{\mathrm{tnn}}_{k,n+k}$}
\label{sec:IR}

In the previous section we showed how our string construction explains
the classification of possible IR fixed points by cells of Gr$^{\mathrm{tnn}}_{k,n+k}$.
In this section we show that the coordinates of the Grassmannian
are also captured by physical data of our theory. We present
evidence that once we compactify our theory, VEVs of line operators in
the 3D theory provide a canonical basis of coordinates for Gr$^{\mathrm{tnn}}_{k,n+k}$.
The link we establish is based on the mapping of weight assignments on faces of the bipartite
graph to coordinates of Gr$^{\mathrm{tnn}}_{k,n+k}$, which we review in Appendix A. So in other words, it is
enough to give a physical interpretation for the weights of faces in the graph.

The class of objects we consider is dictated by the geometry of the bipartite graph.
For each face $f$, there is a canonical choice of
surface operator given by a D3-brane suspended between NS5-branes which
wraps a two-chain with boundary on the contour of the face and fills two directions of the
4D spacetime. This D3-brane is induced by a gauge field flux in the
central $U(1)_{\rm D5}$ of a gauge group factor for the
D5-brane which fills this same face. Loosely speaking, we
associate the VEV of this surface operator $\mathcal{V}_f$ with a weight
assignment for the face $f$.

To specify the ``VEV of a surface operator'' in more precise terms, we actually need
to compactify our 4D theory. To this end, we replace our 4D non-compact geometry by
$S_{(t)}^{1}\times MC_{q}$ where $MC_{q}$ denotes the Melvin cigar. The
three-dimensional $MC_{q}$ is given by taking a two-dimensional disk $C$ with
a cigar-like metric, and fibering it over an additional $S_{(M)}^{1}$, where
the coordinate $z$ of $C$ transforms by $z\mapsto qz$ for $q=\exp(i\vartheta)$
a complex phase.\footnote{Dimensional reduction of 4D $\scN=1$ theories
described by bipartite networks have been
discussed recently in \cite{Yamazaki:2012cp,Terashima:2012cx}.
Note that the 4D superconformal index discussed there is defined on
$S^1_{(t)}\times S^3$, and the $MC_q$ discussed here can be thought of as
half of an $S^3$ \cite{Cecotti:2011iy}.}
Our surface operator wraps the $S^{1}_{(t)} \times S_{(M)}^1$ factors.
To preserve supersymmetry we must twist the theory by embedding the
$U(1)_{\rm D5}$ in the $U(1)_R$ R-symmetry of
our superconformal theory. In the limit where the radius of $S^{1}_{(t)}$
shrinks to zero size, we can T-dualize this circle to the geometry $\widetilde{S}^{1}_{(t)} \times MC_{q}$,
obtaining a three-dimensional theory in which the original surface operator
is now given by a $U(1)$ holonomy around the circle $S^{1}_{(M)}$.
We propose to identify the weight of a face $w_{f}$ with the VEV of
the surface operator in the limit where the radius $R_{t}$ of $S^{1}_{(t)}$
tends to zero size:
\begin{equation} \label{LIMIT}
w_{f} = \underset{R_{t}\rightarrow0}{\lim}\mathcal{V}_{f}.
\end{equation}
Note that since we are dealing with an $\mathcal{N}=1$
superconformal theory, the resulting 3D theory will preserve
eight real supercharges, four of which will be preserved by the presence of the D3-brane.

To see why the small $S^{1}_{(t)}$ limit is natural and leads
to a correspondence between face weights and VEVs of line operators in the 3D theory,
first observe that very similar operators appear in the related 4D $\mathcal{N} = 2$ generalized $(A_{k-1} , A_{n-1})$
Argyres-Douglas theories. Consider the 4D $\mathcal{N} = 2$ theory obtained from an M5-brane wrapped on the
curve $\Sigma$ and filling $\mathbb{R}^{4}$. We can compactify this system on the geometry
$\widetilde{S}_{(t)}^{1}\times MC_{q}$. As we have already mentioned, each compact face of the
bipartite graph corresponds to a node of the BPS quiver. Though they
are not part of the $\mathcal{N} = 2$ BPS quiver, we can also include the
non-dynamical flavor $U(1)$ factors associated with the $\mathcal{N} = 1$ theory. For each such $U(1)$
factor there is a corresponding holonomy, i.e. a choice of line operator which
encircles the $S_{(M)}^{1}$. These are 1/2 BPS operators and the choice of
$\mathcal{N}=1$ subalgebra is dictated by the R-twisting, i.e. a choice of
embedding the $U(1)$ in the $SU(2)_{R}$ R-symmetry of the $\mathcal{N}=2$
theory \cite{Cecotti:2010fi}. A line operator specifies a BPS state,
corresponding to an M2-brane which ends on the ramification points of $\Sigma$.

\begin{figure}[t!]
\centering{\includegraphics[scale=0.5]{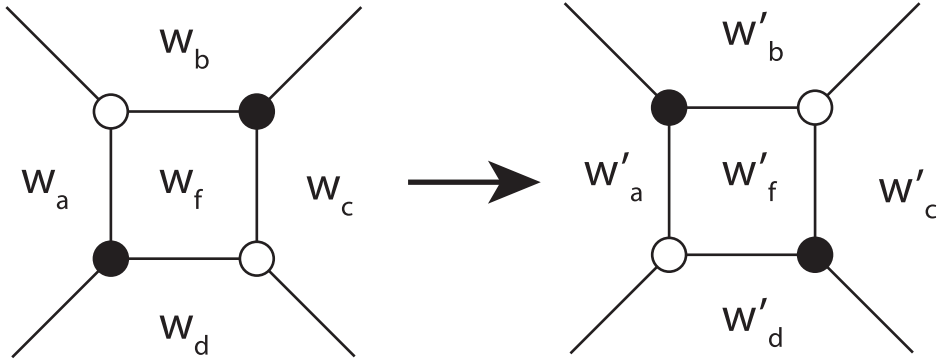}}
\caption{Depiction of the change of the VEVs of line operators in the dimensionally reduced
theory under a Seiberg-like duality. The weights transform as
$w_{f}\rightarrow w'_f=w_{f}^{-1}\ ,\quad w_{a,c}\rightarrow w'_{a,c}=w_{a,c}(1+w_{f}%
)\ ,\quad w_{b,d}\rightarrow w'_{b,d}=w_{b,d}(1+w_{f}^{-1})^{-1}$. This transformation is
precisely what is required for face weights to specify a coordinate system for
Gr$^{\mathrm{tnn}}_{k,n+k}$.}
\label{Seibergweight}
\end{figure}

Now in the $\mathcal{N} = 2$ setting, the corresponding
line operators satisfy the equal time commutator \cite{Cecotti:2010fi}:
\begin{equation}
\mathcal{U}_{\gamma}(t) \mathcal{U}_{\gamma^{\prime}}(t)=q^{\left\langle \gamma,\gamma^{\prime
}\right\rangle } \mathcal{U}_{\gamma^{\prime}}(t) \mathcal{U}_{\gamma}(t) \ ,
\end{equation}
where the pairing $\left\langle \gamma,\gamma^{\prime}\right\rangle $ is the
Dirac-Schwinger-Zwanziger pairing of electric and magnetic states. Moreover, as follows from \cite{Cecotti:2010fi},
under a Seiberg-like duality of the BPS flow lines, each such $\mathcal{U}_{\gamma}$ transforms
according to the weight transformation rules (in the case $q = 1$):
\begin{equation}
w_{f}\rightarrow w'_f=w_{f}^{-1}\ ,\quad w_{a,c}\rightarrow w'_{a,c}=w_{a,c}(1+w_{f}%
)\ ,\quad w_{b,d}\rightarrow w'_{b,d}=w_{b,d}(1+w_{f}^{-1})^{-1} \ ,\label{weighttransf}%
\end{equation}
where the boundary of a face is associated with a given contour $\gamma$. Here, the
$w_{a,b,c,d}$ denote the weight assignments of the faces bordering on
the square which undergoes Seiberg duality (see figure \ref{Seibergweight}). In other words, since surface operators
of the 4D $\mathcal{N} = 1$ theory compactified on $S_{(t)}^{1} \times MC_{q}$ correspond, in the small $S_{(t)}^{1}$ limit,
to line operators of the $\mathcal{N} = 2$ theory compactified on $\widetilde{S}_{(t)}^{1} \times MC_{q}$, we have
determined the transformation rule (\ref{weighttransf}) for
the VEVs $\mathcal{V}_{f}$ in the small radius limit.

Quite remarkably, in the combinatorics literature the transformation of the
face weights in line (\ref{weighttransf}) is precisely what is required in order for these
variables to define coordinates of the totally non-negative Grassmannian
\cite{Postnikov}!\footnote{See \cite{Postnikov} and Appendix A for details. The
boundary measurement map is preserved under the square move if we
simultaneously change the weights following \eqref{weighttransf}.}
In other words, we have shown how the VEVs of line operators associated with the reduction of surface operators
in the 4D theory specify coordinates of the Grassmannian.\footnote{In the $\mathcal{N}=2$ context, this
can be viewed as a statement about the
\textquotedblleft IR\ basis\textquotedblright\ of line operators discussed in \cite{Gaiotto:2010be}.
Alternatively, one can work in terms of a \textquotedblleft
UV\ basis\textquotedblright\ of line operators associated with chords
connecting punctures of the Gaiotto curve. In this alternative basis, these
line operators specify Pl\"{u}cker coordinates of the Grassmannian, which
transform as coordinates of the cluster algebra.
See e.g. \cite{ClusterScott}
for discussion of cluster algebras and coordinates of the Grassmannian in the
math literature. It is helpful to illustrate this construction in the
special case $y^{2}=x^{2}-a^2$, corresponding to a square dimer with four
non-compact faces and one compact internal face. The compact face corresponds
to the single node of the $A_{1}$ BPS quiver, i.e., it has a single $\mathcal{N} = 2$ mass term.
For our purposes, however, we can also include the additional $U(1)$ flavor
symmetries coming from the $\mathcal{N} = 1$ theory. Indeed, this is natural in the context of larger bipartite graph
constructions where these additional faces are automatically compactified. In
this simple setting, we can consider the related Gaiotto curve, which is given
by a $\mathbb{P}^{1}$ with a fourth order pole at one marked point. This can
be represented on the disk by marking four points on its boundary, and drawing
chords between these marked points. The resulting six chords define contours
on the Gaiotto curve, which are in turn mapped to the VEVs of line operators
of the $\mathcal{N}=2$ theory. As found in \cite{Cecotti:2010fi}, these
coordinates form a cluster algebra, and satisfy the Pl\"{u}cker relation (see also
\cite{ClusterScott}):
\begin{equation}
x_{12}x_{34}+x_{14}x_{23}=x_{13}x_{24} \ ,
\end{equation}
where the points on the boundary of the disk are counter-clockwise ordered.
From this identity, we see that there are only five independent line
operators, which matches to the number of face weights of the bipartite
theory. We thank S. Cecotti for helpful discussions on this point.}
Note that the mathematical literature on planar networks \cite{Postnikov}
discusses only the case $q=1$. Our considerations suggest that
there is an even richer $q$-deformation of these results.

\section{Conclusions \label{sec:CONC}}

The stringy realization of a gauge theory frequently provides powerful
geometric insight into the strong coupling behavior of the field theory. In
this paper we have seen that the recently proposed bipartite CFTs of
\cite{Xie:2012mr} admit a stringy realization which explains the seemingly ad
hoc rules found there. The geometry of the bipartite graph corresponds to
one-dimensional flow lines which form the intersection locus in $\mathbb{C}%
^{2}$ between NS5-branes wrapping special Lagrangians and D5-branes wrapped on
an algebraic curve. Varying the parameters of the background geometry with the
asymptotics of the brane configuration held fixed corresponds to Seiberg
duality in the field theory. In addition to providing a stringy explanation for the various
non-trivial field theoretic dualities, we have also presented evidence
that line operators of the 3D theory obtained from compactification on a circle provide
coordinates for Gr$^{\mathrm{tnn}}_{k,n+k}$. In the remainder
of this section we discuss some further avenues of investigation.

In this work we have seen that the $\mathcal{N} = 1$ bipartite theories defined by cells of
Gr$^{\mathrm{tnn}}_{k,n+k}$ exhibit a surprisingly rich phase structure as a function of
$k$ and $n$. It is likely that our string construction, especially in the
large $N$ limit may provide further insights into the IR behavior of these fixed points.
Using our explicit brane construction, it would be quite instructive to construct and study
a gravity dual description.

We have also provided a formal map between the geometry of related objects in
an $\mathcal{N}=2$ theory and a corresponding $\mathcal{N}=1$ superconformal
theory. In particular we have presented a conjectured relationship between the
weights of faces, i.e. the coordinates of Gr$^{\mathrm{tnn}}_{k,n+k}$ and VEVs of line
operators in the 3D dimensionally reduced theory. It would be interesting
to study this proposal further, even for the simple case of the $y^{2}=x^{n}$ theories.

In addition to their appearance in quiver gauge theories realized by D3-branes probing a toric
Calabi-Yau, dimer models on a \textit{torus} have also appeared in the
study of topological strings on the same geometry \cite{Okounkov:2003sp,Ooguri:2009ri}. It is reasonable to ask whether
topological string theory also makes use of planar dimers. Along these lines, it is natural to
conjecture that the basic connection in this context involves the superconformal index
of the $\mathcal{N} = 1$ theories and the topological string partition function, as in
\cite{Vafa:2012fi,Yamazaki:2012cp}. Developing further the details of this correspondence would be
illuminating on both fronts.

Finally, the appearance of bipartite graphs in diverse areas of theoretical
physics is remarkable. In this work we have established a geometric connection
between two such appearances in $\mathcal{N}=2$ and $\mathcal{N}=1$ theories.
Similar bipartite graphs with a very different physical interpretation have
also recently appeared in the study of $\mathcal{N}=4$ Super Yang-Mills
theory \cite{ABCGPT}. It would be exciting to provide a physical link between
these two appearances of planar bipartite graphs.

\section*{Acknowledgments}

We thank the 10th Simons Summer Workshop in Mathematics and Physics, and the Simons Center for Geometry and Physics
for hospitality and providing a stimulating environment where this work was
initiated. We thank A. Neitzke for collaboration at an initial stage of this
work, and for helpful discussions. We also thank N. Arkani-Hamed,
J. Bourjaily, S. Cecotti, D. Morrison and B. Wecht for helpful
discussions. The work of JJH\ is supported by NSF\ grant
PHY-1067976, and that of CV\ is supported by NSF grant PHY-0244821. The work
of DX is supported by DOE grant DE-FG02-90ER40542, and that of MY\ is
supported by the Princeton Center for Theoretical Science.

\appendix
\section{Coordinates of Gr$^{\rm tnn}_{k,n+k}$}\label{sec:measurement}

In this Appendix we explain the mapping between coordinates of the totally non-negative
Grassmannian Gr$^{\rm tnn}_{k,n+k}$ and weights of the bipartite graph.
The physical interpretation of the face weights is given in section \ref{sec:IR}. Our
discussion follows that in \cite{Postnikov}, to which we refer the interested reader
for further details.

\begin{figure}[htbp]
\centering{\includegraphics[scale=0.5]{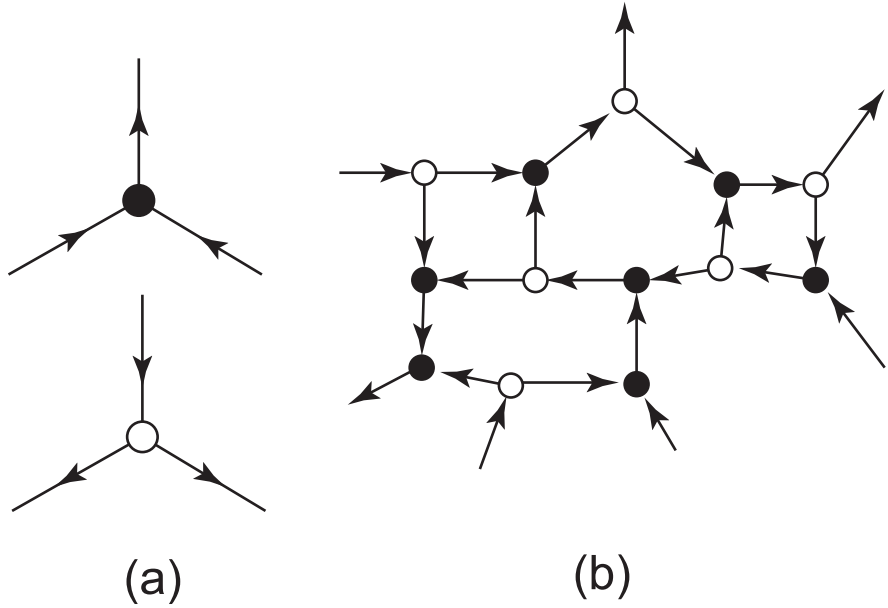}}
\caption{(a) A perfect orientation is a choice of orientation of a bipartite
 graph such that there are two outgoing (resp. incoming) arrows for a black
 (resp. white) vertex. (b) An example of a perfect orientation, with $k=3, n=4$.}
\label{fig.sinksource}
\end{figure}

We first define a perfect orientation.
This is an orientation of edges of bipartite networks
such that each black (white) vertex has one outgoing (incoming) arrow, with the
rest incoming (outgoing), see figure \ref{fig.sinksource} (a).
For a given choice of the perfect orientation, we can consider
the flow defined by the orientation of the arrows inside the
corresponding network. Given an oriented network, $n$ of its
legs are \textquotedblleft sources\textquotedblright%
\ and point into the graph, while $k$ of these are sinks, i.e. attached to
outgoing legs. We can verify that these numbers are the same $k, n$
defined in \eqref{kn}.

We can then define the \textquotedblleft boundary
measurement map\textquotedblright:
\begin{equation}
\text{Meas}:\text{Net}_{k,n+k}\rightarrow {\mathrm{Gr}}^{\rm tnn}_{k,n+k}  \ ,
\label{measurement}
\end{equation}
where $\text{Net}_{k,n+k}$ denotes the set of
planar bipartite networks with given $k$ and $n$. To define this map,
let us assign a positive weight to each edge of the network.
Given a pair of a source and a sink (for a given choice of perfect orientation),
we sum over all the possible oriented paths between them,
whose weight is defined to be
a (signed) product of all the edge weights as we go along the path.
This defines a $k\times n$ matrix. By combining this with a $k\times k$
identity matrix, we can define a $k\times(n+k)$ matrix,
which we define as the image of the boundary measurement map
\eqref{measurement}.

The image of \eqref{measurement} fills out a particular cell specified by the
network, and is independent of the choice of the perfect orientation
chosen in the intermediate stage of the definition.
Also, there is a redundancy in the edge assignments; All that really enters
the matrix are products of weights associated with each face
\cite{Postnikov}. This is the face weight discussed in section
\ref{sec:IR}.\footnote{
In the language of cluster algebras, this is a change from
cluster $x$-variables to cluster $y$-variables (coefficients).
}

\section{Construction of Special Lagrangians}

In section \ref{sec:STRING} we presented a general construction of quiver
gauge theories involving a D5-brane wrapping a holomorphic curve
$\Sigma\subset\mathbb{C}^{2}$ defined by $y^{k}-x^{n}+($deformations$)=0$ and
a class of NS5-branes wrapping special Lagrangian submanifolds. The basic
assumption we made was that given a BPS\ flow line for the curve $\Sigma$,
there exists a corresponding special Lagrangian in $\mathbb{C}^{2}$ which
intersects $\Sigma$ along this flow line. In this Appendix we explain in more
detail why these special Lagrangian submanifolds exist, and detail some of
their properties. After this, we present an explicit construction in the case
$y^{2}=x^{2}-a^{2}$.

\subsection{Existence and Uniqueness}

In this subsection we show that given a BPS\ flow line on the curve $\Sigma$
defined by $y^{k}=x^{n}+($deformations$),$ there is a corresponding special
Lagrangian $L$ which intersects $\Sigma$ along this flow line.

The basic observation is that $\mathbb{R}^{4}$ is hyperkahler, admitting an
entire sphere's worth of complex structures. We can specify a sLag $L$ as an
analytic subvariety with respect to another complex structure, and therefore
give a characterization of possible intersections between $L$ and the original
curve $\Sigma$ in terms of holomorphic geometry of the rotated complex
structure. We begin by introducing the real coordinates $x_{i}$ and $y_{i}$:%
\begin{equation}
x=x_{1}+ix_{2}\text{ and }y=y_{1}+iy_{2} \ .
\end{equation}
In this complex structure, the holomorphic two-form is $\Omega=dx\wedge dy$
and the K\"{a}hler form is $J=\frac{i}{2} \left(  dx\wedge d\overline{x}+dy\wedge
d\overline{y}\right)  $. This is only one of many possible choices of complex
structure. We could alternatively take $\operatorname{Re}e^{-i\theta}%
\Omega=\widetilde{J}= \frac{i}{2} \left(  du\wedge d\overline{u}+dv\wedge d\overline
{v}\right)  $ to be the K\"{a}hler form in a rotated complex structure with
holomorphic coordinates $u$ and $v$:%
\begin{align}
u  &  =\operatorname{Re}x+i\operatorname{Re}e^{-i\theta}y=x_{1}+i\left(
\cos\theta y_{1}+\sin\theta y_{2}\right) \ , \\
v  &  =\operatorname{Im}x-i\operatorname{Im}e^{-i\theta}y=x_{2}+i\left(
\sin\theta y_{1}-\cos\theta y_{2}\right) \ .
\end{align}
Our candidate sLags are therefore given by the zero set of an analytic
function $g(u,v)$:%
\begin{equation}
L=\left\{  g(u,v)=0\right\}
\end{equation}
in the rotated complex structure.

This coordinate system is quite helpful in describing the special Lagrangians
which intersect $y^{k}=x^{n}$ along BPS flow lines. Based on the BPS\ flow
equation $ydx=\alpha dt$, we see that since the $t$ coordinate is a real
parameter, we must demand:%
\begin{equation}
\operatorname{Im} \left[e^{i\theta}x^{(n+k)/k}\right]=0 \ .
\end{equation}
In other words, on one of the sheets of the cover for $\Sigma$, we have:%
\begin{align}
x  &  =e^{-ik\theta/(n+k)}\xi_{n+k}r_{x} \ ,\\
y  &  =e^{-in\theta/(n+k)}\xi_{n+k}^{-1}r_{y} \ .%
\end{align}
For $r_{x}$ and $r_{y}$ real with $\xi_{n+k}$ an $(n+k)^{\rm th}$ root of unity.
The other sheets are obtained by multiplying $y$ by a $k^{\rm th}$ root of unity.

Without loss of generality, we can set $\theta=0$, and work with respect to
the same sheet used above. In terms of the coordinates $u$ and $v$, we have:%
\begin{align}
u  &  =\operatorname{Re}x+i\operatorname{Re}y=x_{1}+iy_{1} \ ,\\
v  &  =\operatorname{Im}x-i\operatorname{Im}y=x_{2}-iy_{2} \ .%
\end{align}
Writing $\xi_{n+k}=\cos\vartheta+i\sin\vartheta$, we therefore have:%
\begin{align}
u  &  =\cos\vartheta r_{x}+i\cos\vartheta r_{y} \ ,\\
v  &  =\sin\vartheta r_{x}+i\sin\vartheta r_{y}\ .
\end{align}
In other words the sLag is specified by:%
\begin{equation}
\frac{u}{v}=\cot\vartheta \ .
\end{equation}
The corresponding algebraic equation describing the sLag is:
\begin{equation}
g(u,v)=\underset{j=1}{\overset{n+k}{%
{\displaystyle\prod}
}}\left(  u\sin\frac{2\pi j}{n+k}-v\cos\frac{2\pi j}{n+k}\right) \ .
\end{equation}

Our discussion so far has involved the explicit form of the sLag in the limit
where all lower order deformations of $\Sigma$ have been switched off. Once we
switch on deformations of $\Sigma$, the explicit parameterization of the sLag
will be quite complicated. Nevertheless, it is still possible to argue for the
existence of a sLag given a particular BPS\ flow line. We work away from
points of ramification of $\Sigma$. In this case, the BPS flows are
characterized by a one-dimensional path on $\Sigma$.

To construct the profile of the sLag in a local patch surrounding the
BPS\ flow line, we proceed as follows. Consider first the trajectory of the
curve in $\mathbb{R}^{4}$ viewed as $\mathbb{C}^{2}$ with respect to the
rotated complex structure $u(t)$, $v(t)$:%
\begin{align}
\gamma &  :\mathbb{R}\rightarrow\mathbb{R}^{4} \ ,\\
t  &  \mapsto(u_{1}(t),u_{2}(t),v_{1}(t),v_{2}(t)) \ .
\end{align}
To extend this curve to a special Lagrangian, we locally \textquotedblleft
fatten\textquotedblright\ this one-dimensional space to a two-dimensional
manifold. In other words, having specified a trajectory, we need to pick one
direction normal to the curve such that the resulting sLag is locally fixed.
The tangent vector along the curve is given by the $\mathbb{R}^{4}$ vector
$(u_{1}^{\prime}(t),u_{2}^{\prime}(t),v_{1}^{\prime}(t),v_{2}^{\prime}(t))$.
The normal directions sweep out an $\mathbb{R}^{3}$ within this space. Out of
these three directions, the one corresponding to the normal direction is the
one compatible with the complex structure specified by $u$ and $v$. This
choice is unique, and is obtained by acting via $\widetilde{J},$ the complex
structure in the rotated basis. More precisely, there is a group of $U(2)$
transformations on the coordinates $(u^{\prime}(t),v^{\prime}(t))$ which leave
the complex structure unchanged. Of this group of $U(2)$ transformations, only
a $U(1)$ subgroup also leaves the holomorphic two-form invariant. The
holomorphic two-form and K\"{a}hler form in the rotated complex structure are
invariant under the rephasing:%
\begin{equation}
(u^{\prime},v^{\prime})\rightarrow(e^{i\phi}u^{\prime},e^{-i\phi}v^{\prime})
 \ .
\end{equation}
Hence, this circle action locally produces a two-manifold which is special
Lagrangian. Thus, in addition to the time coordinate $t$, we can also
introduce a fixed normal direction $n$. The coordinates $u$ and $v$ are now
functions of two coordinates $t$ and $n$. Let us note in passing that this
local of construction of special Lagrangians works in higher dimensions as
well.\footnote{See e.g. \cite{JOYCEI, JOYCEII} and \cite{Joyce:2001xt} for a
review. Essentially quoting from \cite{Joyce:2001xt},
given a real analytic $(m-1)-$ dimensional submanifold $P$ of $\mathbb{C}^{m}$
with a non-vanishing section of $\Lambda^{m-1}TP$ which we denote by $\chi$,
we can construct a one-parameter family of maps $\phi_{t}:P\rightarrow\mathbb{C}^{m}$,
and a corresponding special Lagrangian satisfying the ODE:
\par
$\left(  \frac{d\phi_{t}}{dt}\right)  ^{b}=\left(  \phi_{t}\right)  _{\ast
}\left(  \chi\right)  ^{a_{1}...a_{m-1}}\left(  \operatorname{Re}%
\Omega\right)  _{a_{1}...a_{m}}g^{a_{m}b}$. Locally, this yields a special
Lagrangian parameterized as $L=\left\{  \phi_{t}(p)\text{ for }t\in%
\mathbb{R} \text{ and }p\in P\right\}  $.}

In other words, once we find a special Lagrangian passing through the
prescribed 1-cycle, there is no remaining choice; it is the unique option
available. Locally, we have also seen that the existence of a 1-cycle does not
lead to any obstruction to the construction of the corresponding special
Lagrangian, so we conclude that at least locally, our special Lagrangians
exists. Finally, even though near the curve $\Sigma$ the components of the
sLag are smooth, it could indeed happen that far from the holomorphic curve,
the behavior of the sLag looks singular. This is not really an issue for our
present purposes since we are only interested in the local behavior of the
sLag near $\Sigma$.

\subsection{Explicit Example}

Having given a general local construction of such sLags, it is helpful to
consider an explicit example. We consider $\Sigma$ given by:%
\begin{equation}
y^{2}=x^{2}-a^{2}
\end{equation}
for $a>0$. In this section we construct the sLags which intersect $\Sigma$ along the
BPS flow lines of $\Sigma$, which are specified by the constant
phase trajectories of $ydx = \alpha dt$ for $\alpha$ a complex phase. Throughout, we restrict attention to the case
$\alpha = 1$, so that the phases of $a$ and $\alpha$ are aligned.

A helpful physical model to keep in mind is complexified quantum mechanics with momentum $y$ and position $x$ (see
footnote \ref{footnote:QM}). Studying the BPS flow lines amounts to working in the semi-classical approximation for
this system. To integrate $ydx$, it is helpful to work in terms of the coordinates $X$ and
$Y$ specified as:%
\begin{align}
x &  =a\cosh X \ , \\
y &  =a\sinh Y \ .
\end{align}
This is a non-singular coordinate transformation near the origin. At large
coordinate values, it develops some singularities. However, we view the
complexified phase space as a region in $\mathbb{C}^{2}$, i.e., we delete the
essential singularities in $X$ and $Y$ space.

In terms of these coordinates, the holomorphic curve reads as:%
\begin{equation}
\sinh^{2}Y=\sinh^{2}X \ .
\end{equation}
In other words, we have:%
\begin{equation}
Y=\pm X \label{Yrel}%
\end{equation}
for the two sheets of the Riemann surface.

Next, we note that in this coordinate system, the WKB\ phase also simplifies:%
\begin{equation}
\int ydx=\frac{a^{2}}{4}\left(  \sinh2X-2X\right)  .
\end{equation}
Since we are assuming $a>0$ writing out $X=X_{1}+iX_{2}$, the condition that
$\int ydx$ is real can be written as:%
\begin{equation}
\cosh\left(  2X_{1}\right)  \sin\left(  2X_{2}\right)  =2X_{2} \ .
\end{equation}
This yields two branches. Either we have $X_{2}=0$ or, if $X_{2}$ does not
vanish, we obtain a more complicated trajectory. We are interested in the
branch which intersects $X_{2}=0$, as this is where a consistent flow can
terminate. We can also identify the location of the other flow lines. This
corresponds to taking $\sin X_{2}\simeq1$ and $X_{1}\rightarrow\infty$. In
this limit, we approach an asymptotic trajectory where $\operatorname{Re}%
x\rightarrow0$ and $\operatorname{Im}x\rightarrow\infty$.

This can be viewed as the analytic continuation of the inverted harmonic
oscillator in the complex $x$-plane. A particle moving along this path is
following a steepest descent along which the phase of the path integral can
remain constant. Having solved for the trajectory in the $X$ coordinates, it
follows that we can also deduce the flow in the $Y$ coordinates, via equation
(\ref{Yrel}). The appropriate trajectory on the top sheet of the Riemann
surface is therefore satisfied by the equations:%
\begin{equation}
\cosh\left(  2X_{1}\right)  \sin\left(  2X_{2}\right)  =2X_{2}\text{, }%
Y_{1}=X_{1}\text{, \ and }Y_{2}=X_{2} \ . \label{firstsol}%
\end{equation}

Let us now turn to a parametrization of the sLag in terms of our $u$ and $v$
coordinates. Our aim will be to find an analytic expression in the $u$ and $v$
coordinates which intersects the holomorphic curve along the specified
BPS\ flow lines. Our strategy for doing this will be to consider an explicit
presentation for $u$ and $v$ as functions of the coordinates $X_{1}$ and
$X_{2}$. Using our relation between $X_{1}$ and $X_{2}$, we will aim to
convert this into an analytic relation between $u$ and $v$. To do this, we
view $u(X_{1},X_{2})$ and $v(X_{1},X_{2})$ as functions of two complex
variables $X_{1}$ and $X_{2}$, i.e., we analytically continue the relation of
equation (\ref{firstsol}) to $\mathbb{C}^{2}$.

To this end, let us begin by first writing out in terms of $a$, $X_{1}$ and
$X_{2}$, the form of the coordinates $u$ and $v$ in the rotated complex
structure:%
\begin{align}
u &  =\operatorname{Re}x+i\operatorname{Re}y=a\left(  \cos X_{2}\cosh
X_{1}+i\cos X_{2}\sinh X_{1}\right)  \ , \\
v &  =\operatorname{Im}x-i\operatorname{Im}y=a\left(  \sin X_{2}\sinh
X_{1}-i\sin X_{2}\cosh X_{1}\right)  \ .
\end{align}
It is helpful to consider the analytic expressions:%
\begin{equation}
\frac{iv}{u}=\tan X_{2} , \quad u^{2}-v^{2}=a^{2}\left(  1+i\sinh
2X_{1}\right)  \ .\label{uvrelations}%
\end{equation}
Eliminating the appearance of $X_{1}$ and $X_{2}$, we obtain the equation for
the sLag:%
\begin{equation}
g_{\gamma}(u,v)=0 \ ,\label{sLag}%
\end{equation}
where
\begin{equation}
g_{\gamma}(u,v)\equiv v - u\tanh\left(  \frac{uv}{a^{2}}\sqrt{\frac{v^{2}%
-u^{2}+2a^{2}}{u^{2}-v^{2}}}\right) \ .
\end{equation}
As we originally anticipated, the deformation involves a transcendental
function. This means that there will actually be infinitely many branches of
this function, and so the sLag will actually intersect the holomorphic curve
$\Sigma$ at infinitely many lines. In other words, in addition to the original
BPS\ flow lines, we have an infinite set of additional intersections between
our holomorphic curve and our sLag.

Physically, however, this further subdivision of the $x$-plane is meaningless.
Indeed, these regions are identified with a single gauge group in the 4D field
theory. In order to have treated these regions as defining distinct gauge
group factors, we would need to have a field theory mode, i.e. a light string
in the theory which could condense to lift this region away from the rest. For
the regions of the original bipartite graph, we can see that such modes exist and
are the bifundamentals between the flavor branes. However, we also see that
there is no consistent way to assign a collection of bifundamentals to the
other regions. Indeed, if we had tried to assign a different gauge group
factor to each region we would have obtained an anomalous gauge group factor.


\newpage

\bibliographystyle{titleutphys}
\bibliography{PlanarDimers}

\providecommand{\href}[2]{#2}\begingroup\raggedright\begin{thebibliography}{10}

\bibitem{Xie:2012mr}
D.~Xie and M.~Yamazaki, ``{Network and Seiberg Duality},''
\href{http://arxiv.org/abs/1207.0811}{{\ttfamily arXiv:1207.0811 [hep-th]}}.

\bibitem{Hanany:2005ve}
A.~Hanany and K.~D. Kennaway, ``{Dimer models and toric diagrams},''
\href{http://arxiv.org/abs/hep-th/0503149}{{\ttfamily arXiv:hep-th/0503149
  [hep-th]}}.

\bibitem{Franco:2005rj}
S.~Franco, A.~Hanany, K.~D. Kennaway, D.~Vegh, and B.~Wecht, ``{Brane dimers
  and quiver gauge theories},''
  \href{http://dx.doi.org/10.1088/1126-6708/2006/01/096}{{\em JHEP} {\bfseries
  0601} (2006) 096},
\href{http://arxiv.org/abs/hep-th/0504110}{{\ttfamily arXiv:hep-th/0504110
  [hep-th]}}.

\bibitem{Franco:2005sm}
S.~Franco, A.~Hanany, D.~Martelli, J.~Sparks, D.~Vegh, {\em et al.}, ``{Gauge
  theories from toric geometry and brane tilings},''
  \href{http://dx.doi.org/10.1088/1126-6708/2006/01/128}{{\em JHEP} {\bfseries
  0601} (2006) 128},
\href{http://arxiv.org/abs/hep-th/0505211}{{\ttfamily arXiv:hep-th/0505211
  [hep-th]}}.

\bibitem{Feng:2005gw}
B.~Feng, Y.-H. He, K.~D. Kennaway, and C.~Vafa, ``{Dimer Models from Mirror
  Symmetry and Quivering Amoebae},'' {\em Adv. Theor. Math. Phys.} {\bfseries
  12} (2008) 489--545,
\href{http://arxiv.org/abs/hep-th/0511287}{{\ttfamily arXiv:hep-th/0511287
  [hep-th]}}.

\bibitem{Kennaway:2007tq}
K.~D. Kennaway, ``{Brane Tilings},''
  \href{http://dx.doi.org/10.1142/S0217751X07036877}{{\em Int.J.Mod.Phys.}
  {\bfseries A22} (2007) 2977--3038},
\href{http://arxiv.org/abs/0706.1660}{{\ttfamily arXiv:0706.1660 [hep-th]}}.

\bibitem{Yamazaki:2008bt}
M.~Yamazaki, ``{Brane Tilings and Their Applications},''
  \href{http://dx.doi.org/10.1002/prop.200810536}{{\em Fortsch.Phys.}
  {\bfseries 56} (2008) 555--686},
\href{http://arxiv.org/abs/0803.4474}{{\ttfamily arXiv:0803.4474 [hep-th]}}.

\bibitem{Postnikov}
A.~Postnikov, ``{Total Positivity, Grassmannians, and Networks},''
  \href{http://arxiv.org/abs/0609764}{{\ttfamily arXiv:0609764 [math]}}.

\bibitem{Franco:2012mm}
S.~Franco, ``{Bipartite Field Theories: from D-Brane Probes to Scattering
  Amplitudes},''
\href{http://arxiv.org/abs/1207.0807}{{\ttfamily arXiv:1207.0807 [hep-th]}}.

\bibitem{Ooguri:2008yb}
H.~Ooguri and M.~Yamazaki, ``{Crystal Melting and Toric Calabi-Yau
  Manifolds},'' \href{http://dx.doi.org/10.1007/s00220-009-0836-y}{{\em
  Commun.Math.Phys.} {\bfseries 292} (2009) 179--199},
\href{http://arxiv.org/abs/0811.2801}{{\ttfamily arXiv:0811.2801 [hep-th]}}.

\bibitem{Intriligator:2003jj}
K.~A. Intriligator and B.~Wecht, ``{The Exact superconformal R symmetry
  maximizes a},'' \href{http://dx.doi.org/10.1016/S0550-3213(03)00459-0}{{\em
  Nucl. Phys.} {\bfseries B667} (2003) 183--200},
\href{http://arxiv.org/abs/hep-th/0304128}{{\ttfamily arXiv:hep-th/0304128
  [hep-th]}}.

\bibitem{Kutasov:1995np}
D.~Kutasov and A.~Schwimmer, ``{On duality in supersymmetric Yang-Mills
  theory},'' \href{http://dx.doi.org/10.1016/0370-2693(95)00676-C}{{\em
  Phys.Lett.} {\bfseries B354} (1995) 315--321},
\href{http://arxiv.org/abs/hep-th/9505004}{{\ttfamily arXiv:hep-th/9505004
  [hep-th]}}.

\bibitem{Kutasov:2003iy}
D.~Kutasov, A.~Parnachev, and D.~A. Sahakyan, ``{Central charges and U(1)(R)
  symmetries in N=1 superYang-Mills},'' {\em JHEP} {\bfseries 0311} (2003) 013,
\href{http://arxiv.org/abs/hep-th/0308071}{{\ttfamily arXiv:hep-th/0308071
  [hep-th]}}.

\bibitem{Intriligator:2003mi}
K.~A. Intriligator and B.~Wecht, ``{RG fixed points and flows in SQCD with
  adjoints},'' \href{http://dx.doi.org/10.1016/j.nuclphysb.2003.10.033}{{\em
  Nucl.Phys.} {\bfseries B677} (2004) 223--272},
\href{http://arxiv.org/abs/hep-th/0309201}{{\ttfamily arXiv:hep-th/0309201
  [hep-th]}}.

\bibitem{Cecotti:2010fi}
S.~Cecotti, A.~Neitzke, and C.~Vafa, ``{R-Twisting and 4d/2d
  Correspondences},''
\href{http://arxiv.org/abs/1006.3435}{{\ttfamily arXiv:1006.3435 [hep-th]}}.

\bibitem{Alim:2011ae}
M.~Alim, S.~Cecotti, C.~Cordova, S.~Espahbodi, A.~Rastogi, {\em et al.}, ``{BPS
  Quivers and Spectra of Complete N=2 Quantum Field Theories},''
\href{http://arxiv.org/abs/1109.4941}{{\ttfamily arXiv:1109.4941 [hep-th]}}.

\bibitem{Xie:2012dw}
D.~Xie, ``{Network, Cluster coordinates and N=2 theory I},''
\href{http://arxiv.org/abs/1203.4573}{{\ttfamily arXiv:1203.4573 [hep-th]}}.

\bibitem{Gaiotto:2012rg}
D.~Gaiotto, G.~W. Moore, and A.~Neitzke, ``{Spectral networks},''
\href{http://arxiv.org/abs/1204.4824}{{\ttfamily arXiv:1204.4824 [hep-th]}}.

\bibitem{Gaiotto:2012db}
D.~Gaiotto, G.~W. Moore, and A.~Neitzke, ``{Spectral Networks and Snakes},''
\href{http://arxiv.org/abs/1209.0866}{{\ttfamily arXiv:1209.0866 [hep-th]}}.

\bibitem{Shapere:1999xr}
A.~D. Shapere and C.~Vafa, ``{BPS structure of Argyres-Douglas superconformal
  theories},''
\href{http://arxiv.org/abs/hep-th/9910182}{{\ttfamily arXiv:hep-th/9910182
  [hep-th]}}.

\bibitem{Ooguri:1996me}
H.~Ooguri and C.~Vafa, ``{Summing up D-Instantons},''
  \href{http://dx.doi.org/10.1103/PhysRevLett.77.3296}{{\em Phys. Rev. Lett.}
  {\bfseries 77} (1996) 3296--3298},
\href{http://arxiv.org/abs/hep-th/9608079}{{\ttfamily arXiv:hep-th/9608079
  [hep-th]}}.

\bibitem{Witten:2010zr}
E.~Witten, ``{A New Look At The Path Integral Of Quantum Mechanics},''
\href{http://arxiv.org/abs/1009.6032}{{\ttfamily arXiv:1009.6032 [hep-th]}}.

\bibitem{Yamazaki:2012cp}
M.~Yamazaki, ``{Quivers, YBE and 3-manifolds},''
  \href{http://dx.doi.org/10.1007/JHEP05(2012)147}{{\em JHEP} {\bfseries 1205}
  (2012) 147},
\href{http://arxiv.org/abs/1203.5784}{{\ttfamily arXiv:1203.5784 [hep-th]}}.

\bibitem{Terashima:2012cx}
Y.~Terashima and M.~Yamazaki, ``{Emergent 3-manifolds from 4d Superconformal
  Indices},'' \href{http://dx.doi.org/10.1103/PhysRevLett.109.091602}{{\em
  Phys.Rev.Lett.} {\bfseries 109} (2012) 091602},
\href{http://arxiv.org/abs/1203.5792}{{\ttfamily arXiv:1203.5792 [hep-th]}}.

\bibitem{Cecotti:2011iy}
S.~Cecotti, C.~Cordova, and C.~Vafa, ``{Braids, Walls, and Mirrors},''
\href{http://arxiv.org/abs/1110.2115}{{\ttfamily arXiv:1110.2115 [hep-th]}}.

\bibitem{Gaiotto:2010be}
D.~Gaiotto, G.~W. Moore, and A.~Neitzke, ``{Framed BPS States},''
\href{http://arxiv.org/abs/1006.0146}{{\ttfamily arXiv:1006.0146 [hep-th]}}.

\bibitem{ClusterScott}
J.~S. Scott, ``{Grassmannians and cluster algebras},'' {\em Proc. London. Math.
  Soc. (3)} {\bfseries 92} (2006) 345--380,
  \href{http://arxiv.org/abs/math/0311148}{{\ttfamily math/0311148}}.

\bibitem{Okounkov:2003sp}
A.~Okounkov, N.~Reshetikhin, and C.~Vafa, ``{Quantum Calabi-Yau and classical
  crystals},'' {\em Progr.Math.} {\bfseries 244} (2006) 597,
\href{http://arxiv.org/abs/hep-th/0309208}{{\ttfamily arXiv:hep-th/0309208
  [hep-th]}}.

\bibitem{Ooguri:2009ri}
H.~Ooguri and M.~Yamazaki, ``{Emergent Calabi-Yau Geometry},''
  \href{http://dx.doi.org/10.1103/PhysRevLett.102.161601}{{\em Phys.Rev.Lett.}
  {\bfseries 102} (2009) 161601},
\href{http://arxiv.org/abs/0902.3996}{{\ttfamily arXiv:0902.3996 [hep-th]}}.

\bibitem{Vafa:2012fi}
C.~Vafa, ``{Supersymmetric Partition Functions and a String Theory in 4
  Dimensions},''
\href{http://arxiv.org/abs/1209.2425}{{\ttfamily arXiv:1209.2425 [hep-th]}}.

\bibitem{ABCGPT}
N.~Arkani-Hamed, J.~L. Bourjaily, F.~Cachazo, A.~B. Goncharov, A.~Postnikov,
  {\em et al.}, ``{Scattering Amplitudes and the Positive Grassmannian},''
\href{http://arxiv.org/abs/1212.5605}{{\ttfamily arXiv:1212.5605 [hep-th]}}.

\bibitem{JOYCEI}
D.~Joyce, ``{Constructing special Lagrangians $m$-folds in $\mathbb{C}^m$ by
  evolving quadrics},'' {\em Math. Ann.} {\bfseries 320} (2001) 757--797,
  \href{http://arxiv.org/abs/math.DG/0008155}{{\ttfamily arXiv:math.DG/0008155
  [math.DG]}}.

\bibitem{JOYCEII}
D.~Joyce, ``{Evolution equations for special Lagrangian $3$-folds in
  $\mathbb{C}^3$},'' {\em Global Anal. Geom.} {\bfseries 20} (2001) 345--403,
  \href{http://arxiv.org/abs/math.DG/0010036}{{\ttfamily arXiv:math.DG/0010036
  [math.DG]}}.

\bibitem{Joyce:2001xt}
D.~Joyce, ``{Lectures on Calabi-Yau and special Lagrangian geometry},''
\href{http://arxiv.org/abs/math/0108088}{{\ttfamily arXiv:math/0108088
  [math-dg]}}.

\end{thebibliography}\endgroup

\end{document}